\documentclass{egpubl}
 
\JournalSubmission    %

\CGFStandardLicense

\usepackage[T1]{fontenc}
\usepackage{dfadobe}  

\biberVersion
\BibtexOrBiblatex
\usepackage[backend=biber,bibstyle=EG,citestyle=alphabetic,backref=true,maxcitenames=1,uniquelist=false]{biblatex} 
\electronicVersion
\PrintedOrElectronic

\ifpdf \usepackage[pdftex]{graphicx} \pdfcompresslevel=9
\else \usepackage[dvips]{graphicx} \fi

\usepackage{egweblnk} 

\usepackage{epsfig}
\usepackage{graphicx}
\usepackage{amsmath}
\usepackage{float}
\usepackage{xcolor}
\usepackage{array}
\usepackage{wrapfig}
\usepackage{wasysym}
\usepackage{csvsimple}
\usepackage{adjustbox}
\usepackage[percent]{overpic}
\usepackage{pifont}
\usepackage{comment}
\usepackage{hyperref}
\usepackage{amsmath}
\usepackage{graphicx}
\usepackage{kbordermatrix}

\definecolor{amber}{rgb}{1.0, 0.75, 0.0}
\definecolor{applegreen}{rgb}{0.55, 0.71, 0.0}
\definecolor{darkgoldenrod}{rgb}{0.72, 0.53, 0.04}
\definecolor{firebrick}{rgb}{0.7, 0.13, 0.13}

\newcommand{\ourmethod}{GeoCode}

\usepackage{pgfplots}
\pgfplotsset{compat=newest}
\usepgfplotslibrary{groupplots}
\usepgfplotslibrary{dateplot}

\usepackage[bottom]{footmisc}
\newif\ifdraft
\draftfalse

\ifdraft
\newcommand{\rhc}[1]{{\color{blue}[\textbf{Rana:} #1]}}
\newcommand{\ray}[1]{{\color{purple}[\textbf{Ray:} #1]}}
\newcommand{\opl}[1]{{\color{cyan}[\textbf{Ofek:} #1]}}
\newcommand{\kh}[1]{{\color{orange}[\textbf{Kate:} #1]}}
\newcommand{\itai}[1]{{\color{darkgoldenrod}[\textbf{Itai:} #1]}}
\newcommand{\new}{\color[HTML]{000000}}
\newcommand{\sig}{\color[HTML]{116bd1}}

\newcommand{\rh}[1]{{\color{blue}#1}}

\else
\newcommand{\rhc}[1]{}
\newcommand{\ray}[1]{}
\newcommand{\opl}[1]{}
\newcommand{\kh}[1]{}
\newcommand{\itai}[1]{}
\newcommand{\rh}[1]{{\color{black}#1}}

\newcommand{\new}{\color[HTML]{000000}}
\newcommand{\sig}{\color[HTML]{000000}}

\newcommand{\cgf}{\color[HTML]{000000}}
\newcommand{\reva}[1]{{\color[HTML]{000000}#1}}
\fi

\newcommand{\tt}{\texttt}
\newcommand{\bf}{\textbf}

\newcommand{\citeyearbrackets}[1]{[\citeyear{#1}]}

\usepackage[ruled,vlined]{algorithm2e}

\SetCommentSty{mycommfont}

\usepackage{bbold}
\usepackage{makecell}

\def\hlinewd#1{%
\noalign{\ifnum0=`}\fi\hrule \@height #1 %
\futurelet\reserved@a\@xhline} 

\usepackage{lipsum}

\usepackage{booktabs}
\usepackage{makecell}
\usepackage{multirow}
\usepackage{bigdelim}

\usepackage{hep-text}

\usepackage[capitalize]{cleveref}
\crefname{section}{Sec.}{Secs.}
\Crefname{section}{Section}{Sections}
\Crefname{table}{Table}{Tables}
\crefname{table}{Tab.}{Tabs.}

\title{GeoCode: Interpretable Shape Programs}

\author{
Ofek Pearl$^{1,2}$ \quad \quad Itai Lang$^{2}$ \quad \quad Yuhua Hu$^{2}$ \quad \quad Raymond A. Yeh$^{3}$ \quad \quad Rana Hanocka$^{2}$ \\
$^{1}$Tel Aviv University \quad \quad $^{2}$University of Chicago \quad \quad $^{3}$Purdue University \\
{\tt\small ofekpearl@mail.tau.ac.il \quad \{itailang, katehu, ranahanocka\}@uchicago.edu \quad rayyeh@purdue.edu} \\
}

\author[O. Pearl et al.]
{\parbox{\textwidth}{\centering O. Pearl$^{1,2}$\orcid{0009-0001-7266-0441}
        \quad \quad I. Lang$^{2}$\orcid{0000-0003-4066-4293}
        \quad \quad Y. Hu$^{2}$\orcid{0009-0006-5329-696X}
        \quad \quad R.\,A. Yeh$^{3}$\orcid{0000-0003-4375-0680}
        \quad \quad R. Hanocka$^{2}$\orcid{0000-0003-3214-3703} 
        }
        \\
{\parbox{\textwidth}{\centering $^{1}$Tel Aviv University \quad \quad $^{2}$University of Chicago \quad \quad $^{3}$Purdue University \\
{\tt\small ofekpearl@mail.tau.ac.il \quad \{itailang, katehu, ranahanocka\}@uchicago.edu \quad rayyeh@purdue.edu} \\
       }
}
}

\begin{document}

\teaser{
    \includegraphics[width=0.99\textwidth]{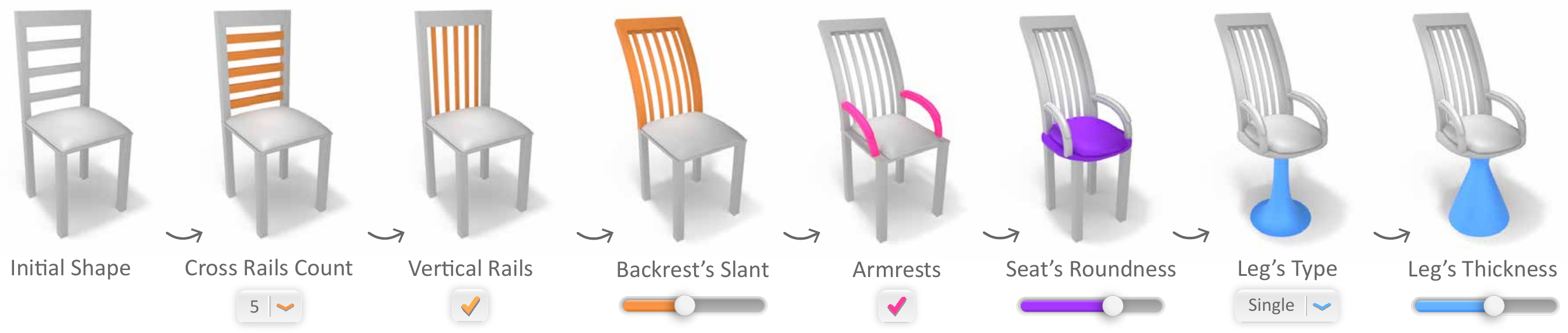}
    \vspace{-0.1cm}
    \caption{\centering\textbf{Interpretable shape control.} GeoCode is a new paradigm that aims to ease the process of creating complex, high-quality procedural shape programs while balancing expressiveness and ease of use. We showcase one of our shape programs, that produces detailed chair shapes while maintaining structural validity through various human-interpretable edits.}
    \label{fig:teaser}
}

\maketitle

\begin{abstract}

{\sig The task of crafting procedural programs capable of generating structurally valid 3D shapes easily and intuitively remains an elusive goal in computer vision and graphics.
Within the graphics community, generating procedural 3D models has shifted to using node graph systems. They allow the artist to create complex shapes and animations through visual programming. Being a high-level design tool, they made procedural 3D modeling more accessible. However, crafting those node graphs demands expertise and training.
We present~\ourmethod{}, a novel framework designed to extend an existing node graph system and significantly lower the bar for the creation of new procedural 3D shape programs. Our approach meticulously balances expressiveness and generalization for part-based shapes. We propose a curated set of new geometric building blocks that are expressive and reusable across domains. We showcase three innovative and expressive programs developed through our technique and geometric building blocks. Our programs enforce intricate rules, empowering users to execute intuitive high-level parameter edits that seamlessly propagate throughout the entire shape at a lower level while maintaining its validity.
To evaluate the user-friendliness of our geometric building blocks among non-experts, we conducted a user study that demonstrates their ease of use and highlights their applicability across diverse domains. Empirical evidence shows the superior accuracy of~\ourmethod{} in inferring and recovering 3D shapes compared to an existing competitor. Furthermore, our method demonstrates superior expressiveness compared to alternatives that utilize coarse primitives. Notably, we illustrate the ability to execute controllable local and global shape manipulations.
Our code, programs, datasets and Blender add-on are available at \mbox{\url{https://github.com/threedle/GeoCode}}.
}


\end{abstract}

\section{Introduction}
\label{sec:intro}
Devising an expressive and intuitive parametrized program that generates structurally valid 3D shapes demands a high level of expertise and is a long-standing goal in computer graphics. A key challenge involves translating the user's edit intent to low-level geometric instructions that will adhere to the desired attributes while maintaining the structural validity of the shape. A promising approach for achieving control over manipulations of 3D shapes is through \emph{procedural methods} that leverage a set of instructions to create a shape.

\begin{figure*}[t!]
    \centering
    \includegraphics[width=0.99\textwidth,height=\textheight,keepaspectratio]{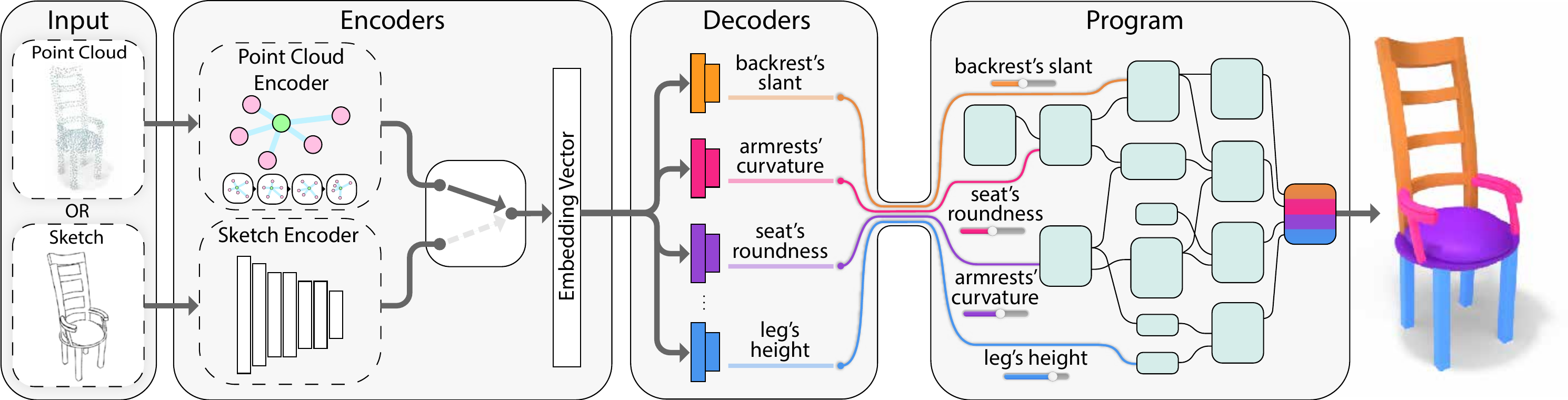}
    \caption{\textbf{System overview.} \ourmethod{} learns to map a point cloud or a sketch input to an intuitively editable parameter space. The input passes through the corresponding encoder to obtain an embedding vector which is then fed to a set of decoders that predict the interpretable parameters. The program enforces a set of rules that, given a parameter representation, produces a high-quality shape by construction. }
    \label{fig:overview}
\end{figure*}

In this work, we present \ourmethod{}, a new paradigm for editing complex, high-quality shapes with programs that are expressive (produce detailed shapes) and executable (enforce structural validity). Notably, \ourmethod{} breaks away from a coarse bounding cuboid representation of shape parts. Instead, we develop rich programs that build shapes from the ground up, using sets of B\'eizer curves. \ourmethod{} programs are capable of producing diverse, human-interpretable, and structurally plausible shape programs that employ various rules such as curves, attachment points, beveling, separation of structure from appearance, symmetries, and more. We conducted a user study that demonstrates that people without any 3D modeling experience were able to understand \ourmethod{}'s design methodology and build procedural shape programs that generate high-quality geometries using \ourmethod{}.

\begin{figure}[b]
    \centering
    \newcommand{\seat}{\color[HTML]{7209b7}}
    \includegraphics[width=0.99\linewidth]{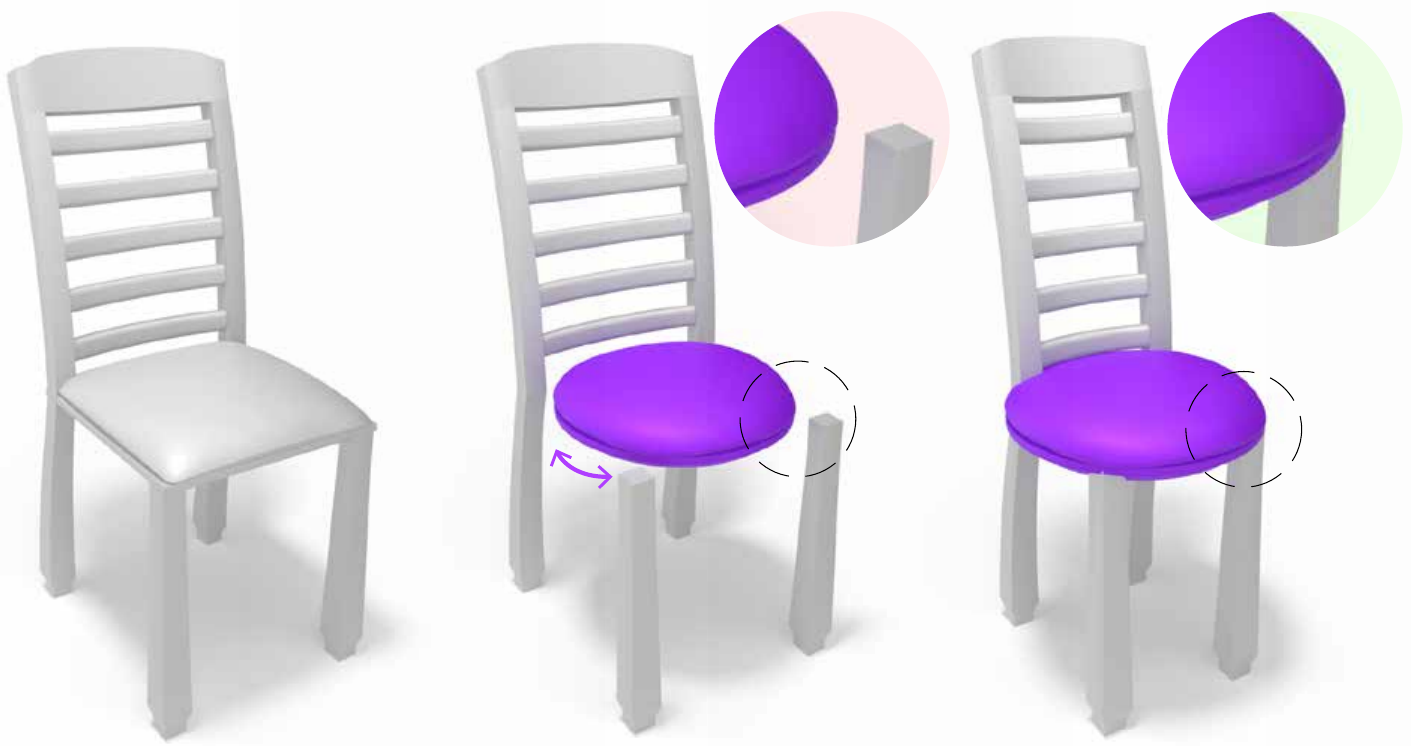}
    \caption{\textbf{Structural integrity.} \ourmethod{} ensures that an edited shape remains structurally plausible. Changing the parameters of the shape (left) to {\seat \textbf{modify the seat}} in isolation will lead to an undesirable result (middle). Our program (right) properly propagates the edit to the remainder of the shape.}
    \label{fig:structural_integrity}
\end{figure}

To demonstrate the effectiveness of our procedural modeling methodology, we built three domain-specific programs. Each program is built for a particular shape domain and exposes a large human-interpretable parameter space, which keeps the shape structurally valid for every combination while capturing geometric features that are represented by the human-interpretable parameter space.
Each parameter space models a wide range of geometric properties and produces a variety of detailed shapes. Shapes can be edited further in an intuitive way and without interacting with the geometry, as shown in \Cref{fig:teaser}. Additionally, the shapes can be easily mixed and interpolated using their interpretable parameter representation and our programs propagate the changes while maintaining a valid shape \rh{(see \Cref{fig:structural_integrity})}.

\rh{
Each program is coupled with a shape recovery network, which learns to infer the program parameters from input point clouds or sketches. We leverage the effectiveness of our procedural programs to generate a large dataset for each of the domains by sweeping over their corresponding parameter space. We train on our automatically generated datasets, to learn a mapping from the input modality (sketch or point cloud) to the interpretable program parameter space, which in turn yields a structured and easily editable shape. 
}
We show that our system generalizes to inputs from different distributions than the training set, such as free-form user-created sketches, sketches generated from images in the wild, noisy point cloud data, real-world point cloud scans, and more.

\reva{In summary, we propose a framework to simplify the process of building general part-based procedural programs. Our contributions are summarized in three main points:
\begin{itemize}
\item \textbf{Blender Add-On:} Our main contribution is allowing non-experts to create expressive procedural 3D programs of part-based shapes. To allow that, we supply a Blender-based \cite{blender} add-on providing users with custom nodes that follow our design methodology. We prove its effectiveness in a user study in which non-expert users built 3D programs for a predefined class (vases) but were also able to build a program for a completely novel class (ceiling lamps) using an illustration that defines the requirements and basic guidance in a timed exercise. This demonstrates that the building blocks in GeoCode are modular enough to extend to novel shape domains even by non-experts.

\item \textbf{Procedural Programs:} We build three domain-specific programs (chairs, vases, and tables) using our procedural modeling methodology, all of which have a large parameter space and offer great variability in the shapes they produce. Another expert program (cabinets) was developed for further discussion and the testing of \ourmethod{}'s modularity.

\item \textbf{Parametric Prediction Network:} We present a system that learns to infer the parameters of the shape program from an input point cloud or sketch. Given a shape program, all the network components are automatically adjusted, from dataset generation to inference. We compare our network with a competing parametric network and present measurable improvements.
\end{itemize}
}
\rh{As far as we can ascertain, we are the first to develop a bottom-up curve-based framework aimed at editing shapes in a structurally valid manner. Given that there are no direct methods for comparison, we compare and evaluate our method to what alternative methods can achieve with their baselines, and demonstrate quantitative and qualitative improvements.}

\section{Related Work}

\smallskip
\noindent \textbf{{\sig Procedural modeling.}}
{\sig Procedural modeling is the process of generating shapes based on given rules. The benefit of using rules to generate shapes is that the input is much more limited than the shape that is produced, and additionally, making modifications or generating variations of the shape is a much less involved process that does not require the artist to remodel the shape.
Contrary to the use of \emph{neural implicit representations} of geometry~\cite{im-net_2018, mescheder2019occupancy, deep-sdf_2019, hao2020dualsdf, sdf-style-gan_2022, hertz2022spaghetti} the resulting procedurally generated shapes can be integrated into any existing 3D pipeline and the artist can freely modify the shape using existing visual graphical user interfaces.}

{\sig Various methods were used to generate the procedural model, an early form of procedural modeling is a grammar used to describe plants known as L-Systems ~\cite{lindenmayer1968mathematical}. It was also successfully applied for buildings ~\cite{muller2006procedural} and for cities~\cite{parish2001procedural}. Another type of grammar is \emph{set grammar} where shapes are considered as symbols~\cite{stiny1982spatial, wonka2003instant}. An extension of set grammar is \emph{shape grammar} which was first introduced in~\cite{stiny1971shape}, where the basic primitives are geometries that may change by the rules (for example, when one geometry is attached to another). This method saw a lot of success in urban design as evident {\cgf from works such as}~\cite{muller2006procedural, schwarz2015advanced, alfadalat2023procedural, lipp2008interactive, willis2021volumetric}.}

{\sig Another class of procedural modeling is \emph{model synthesis} where a 3D model is given as an input example and a larger more complex model is generated~\cite{merrell2007example, merrell2008continuous, merrell2010model}. Many works also provided the users with an interactive way to edit the shapes through the use of a visual system ~\cite{lipp2008interactive, lintermann1999interactive, niese2022procedural, demir2016proceduralization, vanegas2012inverse, vanegas2009interactive, mccrae2008sketch}}. In this work, we explore procedural modeling through the use of visual programming and more specifically with the use of node graph systems. There have been a few works that explored such visual programming frameworks, the work~\cite{barroso2012visual} created such a framework in an attempt to simplify the development process of procedural buildings. Another attempt to create a more generalized framework through the use of a novel node graph system is the work~\cite{ganster2007integrated} and notably, it also allowed some interactive editing in the viewport. Through our user study, we show that our method is intuitive even for non-experts, and with only a little guidance, our participants were able to successfully generate intricate shape programs.

\smallskip
\noindent \textbf{\sig Inverse procedural modeling.}
{\sig The second part of our work resembles inverse procedural modeling in the sense that we predict the input parameters to our procedural models given a point cloud or a sketch of the shape. Similar works include~\cite{stava2014inverse, ritchie2015controlling}, and another work that based the prediction on a neural network~\cite{huang2017shapesynth}. The approach taken by~\cite{huang2017shapesynth} was to split the continuous and discrete parameters into two separate networks. We later use it as a baseline for our work for sketch inputs. However, the work~\cite{huang2017shapesynth} used existing procedural models~\cite{mech2012deco} and accepted only sketches as input. In contrast, in our work, the main goal is to lower the bar for procedural model creation for non-experts. {\cgf A recent work ~\cite{hossain2023data} used inverse procedural modeling to aid expert programmers in iteratively crafting programs based on a given collection of reference shapes. We acknowledge another effort related to the parameter space, where auto-encoders were used to map complex parameter spaces to lower-dimensional ones to ease the process of generating shapes from procedural programs ~\cite{yumer2015procedural}.}

Considering our shape recovery network and programs, a single structure-aware program built using our method and visual programming can encode various structures of the shape and relations between parts from its construction. However, other works tried to solve a more difficult problem, where the relations between parts are also inferred, either as grammar or code ~\cite{shapeAssembly, shapeMOD, wu2013inverse, vanegas2010building, PLAD, merrell2023example, jones2024learning, kapur2024diffusion} or as graphs, often also inferring relations between the parts~\cite{li2017grass, mo2019structurenet, wang22machine, Paschalidou2020CVPR}. 
We compare our work to~\cite{shapeAssembly, mo2019structurenet} solely in terms of the expressibility that our programs can achieve compared to methods that use cuboids as primitive geometry to build the shapes.
There is also limited work on inferring the visual graph node for {\cgf procedural materials} ~\cite{paul2022matformer, hu2022diffproxies, hu2023generating}. {\cgf Some works first retrieve a pre-built procedural material from a database and then optimize its parameters to match a given material ~\cite{hu2019texture, Shi2020match, tchapmi2022generating}.}
We also recognize other works that operate with another procedural modeling technique called constructive solid geometry (CSG) such as InverseCSG~\cite{tao2018inversecsg} or PLAD~\cite{PLAD}.}

\smallskip
\noindent \textbf{Shape reconstruction and editing.}
Several works used deep learning methods to reconstruct a 3D object from a sketch. \textcite{3d-sketch-deep-volumetric} and \textcite{sketch-based-freeform-surface-modeling} predict voxel grids and depth with normal maps, respectively, which are then converted to meshes. Other approaches use Poisson Surface Reconstruction~\cite{kazhdan2006poisson, kazhdan2013poissonrecon} to convert a predicted point cloud to a mesh \cite{lun2017SketchModeling, deepsketch2020, yue2021sketchshapegeneration}, deform a template mesh \cite{zhang2021sketch2model}, learn an implicit representation \cite{Cheng2022crossmodal}, or differentiable mesh representation~\cite{remelli2020meshsdf, guillard2021sketch2mesh}. In contrast, our approach directly outputs a high-quality and editable mesh.

Another approach produces a CAD shape to improve the structural integrity and editing capabilities of the reconstructed mesh. Sketch2CAD \cite{sketch2cad} and Free2CAD \cite{free2cad} trained neural networks to parse 2D sketches into sequences of CAD commands and ComplexGen~\cite{GuoComplexGen2022} created CAD shapes from input point clouds.
However, editing one CAD operation changes only a part of the shape, making the preservation of its structural integrity challenging. In our method, changing one parameter results in local changes that propagate to the rest of the shape, maintaining physical validity while keeping other geometric features intact.

We note that there is a large body of research on recovering the underlying surface mesh from a point cloud input~\cite{hoppe1992surface, kazhdan2006poisson, kazhdan2013poissonrecon, point2mesh, metzer2021orienting}. Our goal in this paper is different. We aim to produce an intuitively editable mesh version of the point cloud, while surface reconstruction works focus mainly on recovering a holistic 3D shape.

\begin{figure*}[t!]
    \centering
    \includegraphics[width=0.99\textwidth]{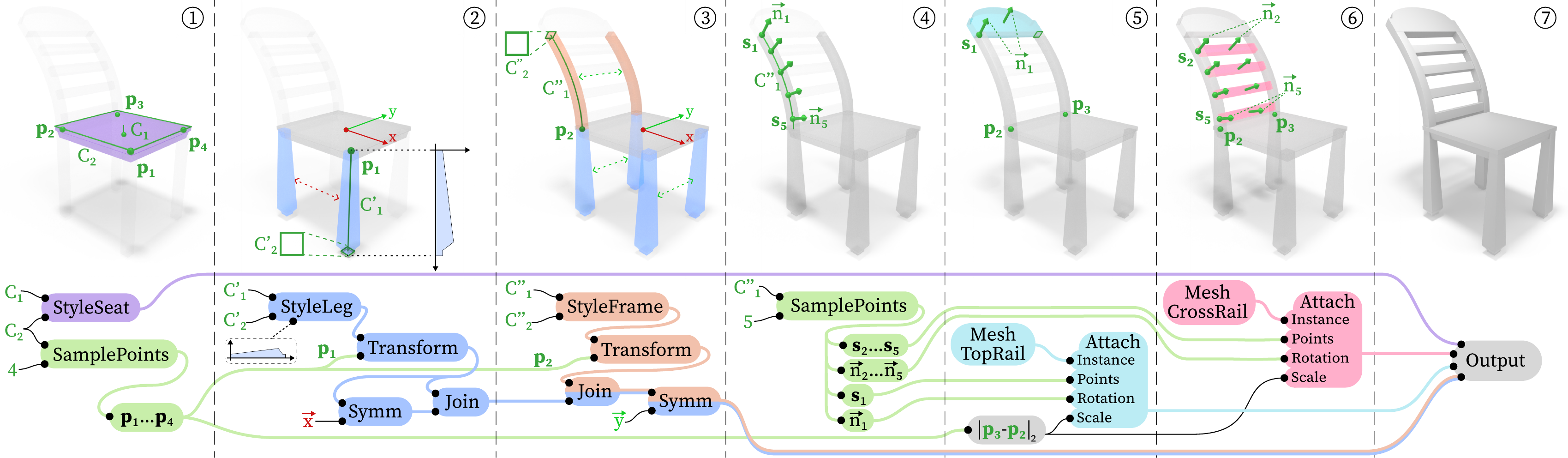}
    \caption{\textbf{Procedural shape construction.} \ourmethod{} generates a set of low-level instructions that adhere to high-level rules to produce a structurally valid shape. We show an example of using \ourmethod{}. In step~{\large\ding{192}}, curve $C_{\tt 1}$, profile curve $C_{\tt 2}$ and the node \textit{StyleSeat} form a triplet that generates the seat. $C_{\tt 2}$ is sampled 4 times which gives us the attachment points $\mathbf{p}_{\tt 1}..\mathbf{p}_{\tt 4}$. The profile curve $C_{\tt 2}$ can parametrically deform into a circle, while the sampled points follow along with those changes. In step~{\large\ding{193}}, curve $C'_{\tt 1}$ and profile curve $C'_{\tt 2}$ generate the leg's mesh when combined with \textit{StyleLeg}. A parameterized function determines the scale of the profile along $C'_{\tt 1}$ \eg~to make the leg thicker at the bottom. The leg is then attached to point $\mathbf{p}_{\tt 1}$. Next, we mirror the leg using a symmetry node \textit{Symm} along the $x$ axis. Lastly, we \textit{Join} the mesh of both legs and pass them further as one unit. In step~{\large\ding{194}}, the frame of the backrest is generated by the triplet of B\'ezier curve $C''_{\tt 1}$ profile curve $C''_{\tt 2}$ and a \textit{StyleFrame} node. The frame mesh is attached to point $\mathbf{p}_2$. We \textit{Join} the new frame and the legs from the previous step and use a \textit{Symm} node to replicate them to the other side of the chair. Step~{\large\ding{195}} shows how attachment points for the top-rail and cross-rails are created dynamically on $C''_{\tt 1}$ by sampling it 5$\times$ using \textit{SamplePoints} node. The sampling node outputs the points and the normal at each point. In step~{\large\ding{196}}, a top-rail shape is attached to the top of the frame at point $\mathbf{s}_{\tt 1}$ and is oriented to match the normal $\bar{\mathbf{n}}_{\tt 1}$. We scale its width to connect seamlessly to both sides of the frame. In step~{\large\ding{197}}, the cross rails are attached at points $\mathbf{s}_{\tt 2}$..$\mathbf{s}_{\tt 5}$ with orientations matching $\bar{\mathbf{n}}_{\tt 2}..\bar{\mathbf{n}}_{\tt 5}$ respectively, and scaled to fit between both sides of the frame. We obtain the final shape (step~{\large\ding{198}}) after joining the parts together.}
    \label{fig:program_inside_look}
\end{figure*}

\section{Method} \label{sec:method}
Our main goal is to simplify the process of building procedural programs, even for complex shapes. We do this, first, by defining a procedural modeling methodology, and secondly, by providing the user with an extension to Blender's Geometry Node graph system~\cite{blender-geometry-nodes} in the form of an easy-to-install add-on. Upon installation, the user will be provided with custom nodes that adhere to our procedural modeling methodology.
Our second goal in this work is to evaluate the expressiveness and effectiveness of programs built using our method. Given an input object, represented as a 3D point cloud or a 2D sketch image, we want to recover the parameter set that will form a 3D shape that best matches the given input. We develop three domain-specific procedural programs parameterized by a very large human-interpretable parameter set. The 3D shape is recovered by training a neural network to infer the set of parameters that will yield the best matching 3D shape once fed to the procedural program; See~\Cref{fig:overview} for an overview.

\subsection{Procedural Modeling Methodology} \label{subsec:procedual_modeling_methodology}

{\sig With the main goal of simplifying the creation of new procedural models through visual programming, we needed to balance two opposing forces. On the one hand, we want to supply non-expert users with nodes that can easily generate complex meshes with multiple input parameters. On the other hand, we want the nodes to be reusable across many shape domains. In other words, we want the nodes we provide to have a lot of expressiveness but we also want them to generalize. Our procedural modeling methodology achieves this by the use of several notions which we explain below: triplets, curve sampling, and symmetries.}

\smallskip
\noindent \textbf{Triplets.}
Guided by that notion, we modeled shape elements using \textit{triplets}. A triplet is a subgraph comprised of three nodes. Firstly, two curves: an open curve $C_{\tt 1}$, which describes a path in the 3D space, and a planar, closed curve $C_{\tt 2}$, which describes the profile of the shape. Secondly, a style node receives the curve and the profile curve as inputs. In the simplest form of the style node, the profile curve $C_{\tt 2}$ is extruded along the curve $C_{\tt 1}$, and creates the 3D mesh.

\begin{figure}[t]
    \centering
    \includegraphics[width=0.90\linewidth]{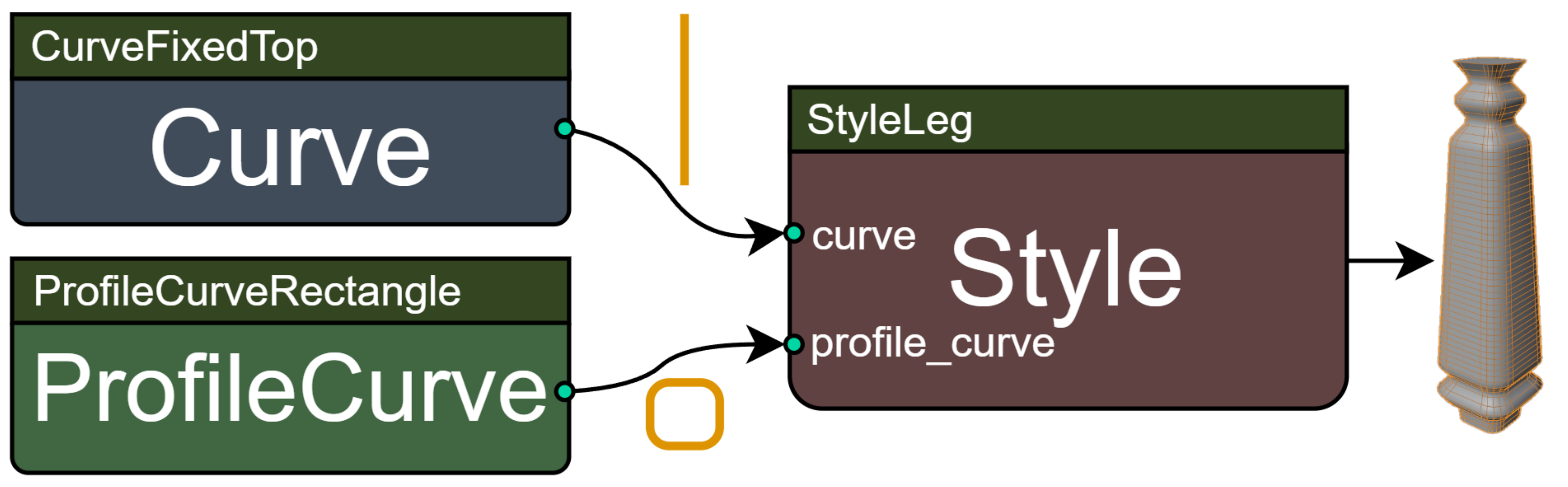}
    \caption{\textbf{Triplets.} Triplets are built using three nodes. The style node accepts a curve node and a profile curve node. It sets the scale at each point along the curve based on given parameters and generates a complex mesh. Here a \ourmethod{} triplet is used to generate a parametric leg with serrations and variable thickness.}
    \label{fig:triplets}
\end{figure}

Within the style node, additional control on the appearance of the final 3D mesh is achieved by setting the scale of the profile $C_{\tt 2}$ at each point along the curve $C_{\tt 1}$. A triplet and its links are demonstrated in \Cref{fig:triplets} where a leg of a chair is generated using only three nodes. Such a leg design may be controlled by 4 continuous parameters: 1) the length of the leg; 2) leg bottom thickness; 3) leg serrations; 4) the shape of the profile curve $C_{\tt 2}$ (\eg interpolate it from a square to a circle). In our add-on design, we make use of curve editor nodes to make the process of adding more styles more intuitive. How a leg triplet might be used within a program is demonstrated in step {\large\ding{193}} in \Cref{fig:program_inside_look}.

\begin{figure*}[t!]
    \centering
    \includegraphics[width=0.99\textwidth]{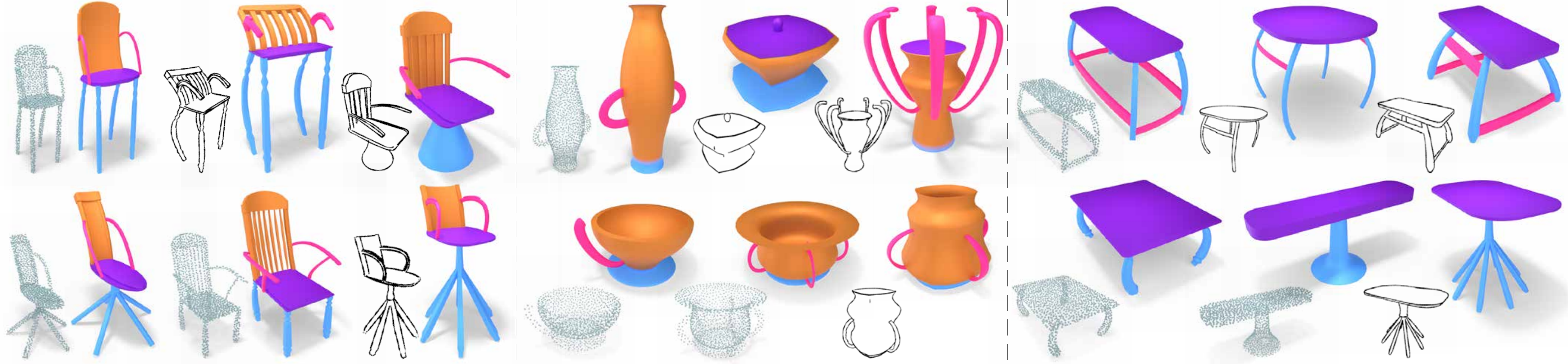}
    \caption{\textbf{Shape gallery.} Showing reconstructed shapes on our test set. Our procedural program produces high-quality geometry given a 3D point cloud or a 2D sketch and contains consistent part segmentation information across the resulting shapes.}
    \label{fig:gallery}
\end{figure*}

We note that on some occasions it makes sense to use mesh primitives such as a sphere or a plane. We use this, for example for the vase's lid handle (see top middle vase in \Cref{fig:gallery}). We can also deform primitives using parameterized mathematical expressions that are applied per vertex.

\smallskip
\noindent \textbf{Curve sampling.}
An artist may approach building shape programs with primitives or pre-modeled parts that are parameterized with blend shapes, a method based on shape interpolation. However, a notable benefit of cultivating curves to build shape elements is the ease of defining attachment points on a curve, by setting points with relative distances along the curve. Meaning, given a shape element $\mathcal{A}$, which is built from a curve $C_{\tt 1}$ and a profile curve $C_{\tt 2}$, an attachment point $\mathbf{p}_{\tt 1}$ on the shape element is defined by a single float number $\mathbf{p}_{\tt 1} \in [0,1]$. A value of $0.0$ is the start point of the curve $C_{\tt 1}$, while a value of $1.0$ is the endpoint of the curve. We attach a shape element $\mathcal{B}$ to shape element $\mathcal{A}$ by defining an attachment point on each one. We optionally set the orientation of $\mathcal{B}$ to match the normal of $C_{\tt 1}$ at the attachment point $\mathbf{p}_{\tt 1}$. Steps {\large\ding{196}} and {\large\ding{197}} in \Cref{fig:program_inside_look} show how the top rail and cross-rails are attached to the frame of the chair in this manner. Scaling shape elements is achieved by calculating distances between attachment points.

\smallskip
\noindent \textbf{Symmetries.}
Other structural relations do not require attachment points and rely simply on symmetry. This is shown in steps {\large\ding{193}} and {\large\ding{194}} in \Cref{fig:program_inside_look}, where we use reflective symmetry to replicate the chair's legs and frame. We also employ rotational symmetry, for example, chairs with swivel legs and vases that have multiple handles as shown in~\Cref{fig:gallery}.

\begin{figure}[t!]
    \centering
    \includegraphics[width=0.93\linewidth]{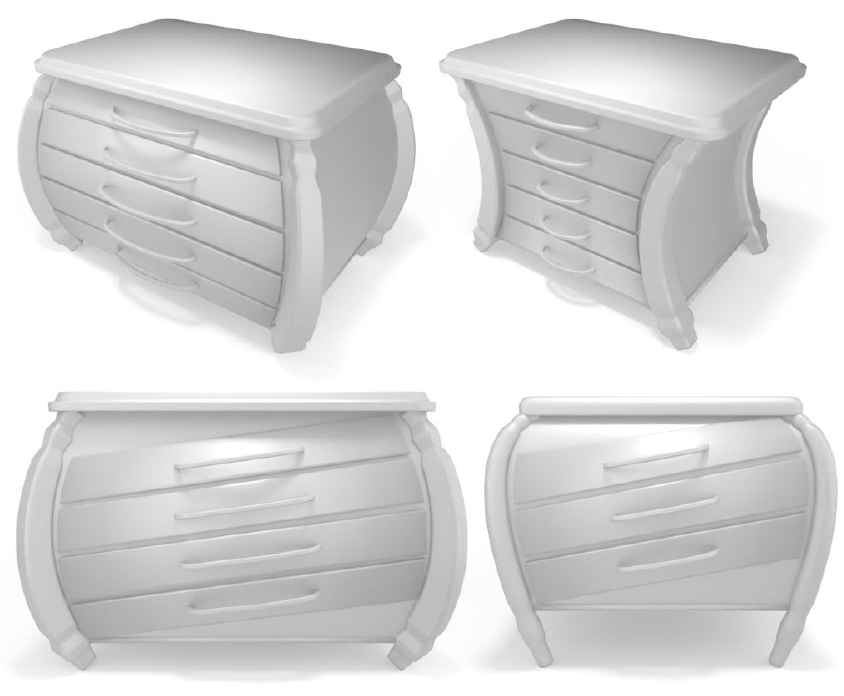}
    \vspace{-0.3cm}
    \caption{\textbf{\reva{Structural integrity discussion.}} \reva{We show that \ourmethod{} can be used in cases where the shape elements depend on two points. The drawers in this example stay between the cabinet frames no matter the width. The handles of the drawers are also nodes in \ourmethod{} add-on and they remain attached even when the drawers are at an angle due to a simulated indexing error.}}
    \label{fig:structural_integrity_discussion}
    \vspace{-0.3cm}
\end{figure}

\noindent \textbf{\reva{Structural integrity discussion.}}
\reva{To invoke a discussion about how structural integrity is maintained using \ourmethod{}, let us consider \Cref{fig:program_inside_look}. In step~{\large\ding{194}}, consider the B\'ezier curve $C''_{\tt 1}$ and assume we now want to make the chair narrower at the top. With the current program as described in \Cref{fig:program_inside_look}, we assume a constant backrest width, so our cross rails (in step~{\large\ding{197}}) will not fit between a narrowing backrest. However, with a more robust design, and using \ourmethod{}'s sampling node and a repeat-zone system (loop), the correct length between the two frames at each point can be calculated and assigned to the individual cross rails. We exemplify this in \Cref{fig:structural_integrity_discussion} where we take a step further and ensure that the sides of each drawer follow exactly the cabinet's profile, even if there is an error in the indexing (see the two bottom examples). This is done inside Blender's repeat zone, by using \ourmethod{}'s curve array-sampling node, Loft node, and a solidify node (all are supplied in the \ourmethod{} add-on). \ourmethod{} also provides curves that attach at two ends such as the handle in \Cref{fig:program_inside_look}, which stays connected correctly even when introduced with indexing errors that cause the drawers to be misaligned. At times when structural integrity is hard to maintain by rules within the program, \ourmethod{} offers a simple way to detect intersections between parts and rule out samples for training based on the detection.}

\subsection{\ourmethod{} Shape Programs} \label{subsec:programs}

We build three procedural programs based on our procedural modeling methodology. The programs are implemented in Blender's~\cite{blender} Geometry Nodes~\cite{blender-geometry-nodes} as a directed acyclic graph (DAG) comprised of \textit{nodes} and \textit{links}. Nodes can hold anything from a single value, all the way to complex geometry. Links can pass along anything from a single value to complex geometry that may include other attributes at the fundamental geometry level (per vertex, per edge, per face, etc.). Examples of operation nodes include math operations or vector operations; mesh primitives, such as cubes or spheres; line primitives such as a B\'ezier curve; and rigid transformation (translation, rotation, and scale). Selected nodes are parameterized by the input parameters, allowing the user to interact with the program and control the resulting shape. The DAG culminates into a single output node which holds the final 3D shape. 

Our programs support three types of human-intuitive input parameters: discrete, binary, and continuous. %
The programs offer disentangled control over the shape, which enables modification of a specific part while keeping all others intact. It also models complex structural interactions, such that one part influences another, and the latter is adapted automatically to preserve part contact and retain the structural integrity of the shape. %
For example, replacing the square seat with a round one in \Cref{fig:structural_integrity} makes the seat narrower, which also changes the leg's position and decreases the width of the backrest accordingly. The programs also hold the part segmentation information as shown in \Cref{fig:gallery}.

The input to our programs is a set of human-interpretable parameters. These can set the structure of the shape (\eg, the height of a chair's seat or the points where the handles will be attached to a vase) or modify shape elements' appearance (\eg, the width of a leg or the roundness of the seat of a chair). Structural edits affect the open curves (\ie, $C_{\tt 1}$) and propagate to other shape elements using the attachment points and symmetries enforced by our program. 

Edits to the appearance of shape elements affect the profile curves (\ie, $C_{\tt 2}$) and can cause structural edits as well. An example of this case is increasing the seat's roundness, which causes the chair to get narrower by imposing structural edits on the shape that bring the legs and the frame of the chair closer together. Considering the opposite direction, edits to the structure of the shape cannot affect the appearance of any of the shape elements. For example, changing the height of the seat of the chair will not affect the seat's roundness or the appearance of the legs. %

\subsection{Mapping to the Program Space} \label{subsec:mapping}
To map a point cloud or sketch input to the human-interpretable parameter representation, we employ an encoder-decoder network architecture. The encoder embeds the input into a global feature vector. Then, we use a set of decoders where each one translates the embedding vector to a single parameter. Together, the final interpretable representation is obtained. Finally, we run the program and recover the 3D shape.~\Cref{fig:overview} illustrates our system design.

\smallskip
\noindent \textbf{Problem formulation.}
We formulate the shape recovery problem as predicting the human-interpretable parameters from a given point cloud or sketch. Let us denote the program parameters as $\{p_i\}$, where each parameter can take $N_i$ discrete values. Continuous program parameters, such as thickness or height, are discretized uniformly over their range. The ground-truth value of the parameter is encoded by a one-hot vector $\mathbf{y}_i \in \{0,1\}^{N_i}$. The ground-truth representation for all the parameters is the concatenation of all $\{\mathbf{y}_i\}$, which we denote as $\mathbf{y} \in \{0,1\}^{\sum_i \! N_i}$. 

The network prediction of the program parameters is as follows:
\begin{align}
\hat{\mathbf{y}}_{\tt{pc}} = \mathbf{D}(\mathbf{E}_{\tt{pc}}(\mathbf{c})), ~~~
\hat{\mathbf{y}}_{\tt{sketch}} = \mathbf{D}(\mathbf{E}_{\tt{sketch}}(\mathbf{s}))
\end{align}

\noindent where $\hat{\mathbf{y}}_{pc}$ and $\hat{\mathbf{y}}_{sketch}$ are the predicted parameters in one-hot representation from the point cloud $\mathbf{c}$ or sketch $\mathbf{s}$, respectively, $\mathbf{E}_{\tt{pc}}$ is the point cloud encoder and $\mathbf{E}_{\tt{sketch}}$ is the sketch encoder, and $\mathbf{D}$ denotes the shared decoders. 

To train the network, we use our program and construct a dataset of the point cloud, sketch, and ground-truth triplets, \ie, $\mathcal{D} = \{(\mathbf{c},\mathbf{s},\mathbf{y})\}$. This process is automated given a program. Then, we train the network with the loss function:
\begin{equation}
    \mathcal{L} = \frac{1}{|\mathcal{D}|}\sum_{(\mathbf{c},\mathbf{s},\mathbf{y}) \in \mathcal{D}} \text{CE}(\hat{\mathbf{y}}_{\tt{pc}},\mathbf{y}) + \text{CE}(\hat{\mathbf{y}}_{\tt{sketch}},\mathbf{y})
\end{equation}

\noindent where \text{CE} denotes the \textit{cross-entropy loss}. We employ the \textit{part existence label} to address parameters that are not represented in the final shape (\eg a chair with no handles). Please refer to \Cref{subsec:sup_dataset} in the supplementary for more information.

\section{Experiments} \label{sec:exp}
We present our user study on GeoCode, as well as qualitative and quantitative evaluations of our method's performance on shape recovery and editing. We demonstrate \ourmethod's ability to recover 3D shapes from point clouds and sketches from a held-out test set from our dataset and from shapes in the wild.
Furthermore, we demonstrate the editing capabilities of our system on the reconstructed shapes, such as modifying part geometry, mixing two shapes, and interpolating between shapes.
{\cgf Finally, we present an ablation study.}
Please refer to the supplementary material for an {\cgf additional} ablation study, additional experiments, and a discussion about failure cases.

\smallskip
\noindent \textbf{Dataset and implementation details.}
{\cgf Each of our three programs handles a different shape domain of chairs, vases, and tables, and has 59, 39, and 36 human-interpretable parameters, respectively.}

To train and evaluate our system, we automatically generated the train, validation, and test datasets using our shape programs. {\cgf Our datasets are lean compared to the vast number of parameter combinations in each shape program. For each possible choice of a single parameter, we only take 30 random shapes for the training set and three shapes for the validation and test sets. We also ensure the parameter choice is taking effect in each random shape}. For each generated 3D shape, we sample 1500 points using Farthest Point Sampling~\cite{eldar1997FPS} and an additional 800 randomly sampled points. We render the 2D sketch images using Blender's~\cite{blender} rendering engine \textit{Freestyle}~\cite{blender-freestyle-engine} from three camera angles while training only on two of them. Sketches are randomly augmented using horizontal flip, stroke dilation, and stroke erosion.

In total, the training set for each domain contains 9570 chairs, 9330 vases, and 6270 tables. The validation and test sets each contain 957 chairs, 933 vases, and 627 tables. For each domain, we ensure that no two shapes in the dataset are the same.

For the point cloud encoder, we use DGCNN~\cite{wang2019dynamic} and employ a VGG architecture~\cite{simonyan2015a_vgg} for the sketch encoder. The decoder utilizes a multi-layer perceptron with three layers for each program parameter. Since each human-interpretable parameter may have a different number of values, the output layer size of each decoder varies according to the number of possible classes of the parameter the decoder is responsible for. Our system ensures that, given a program, the network adjusts those attributes automatically.

\smallskip
\noindent \textbf{Blender Add-On.}
Blender~\cite{blender} offers a node graph system for procedural modeling called Geometry Nodes~\cite{blender-geometry-nodes}. We built an add-on that extends it by providing a multitude of custom nodes that adhere to the procedural modeling methodology (see~\Cref{subsec:procedual_modeling_methodology}). For a list of the nodes we provide, please refer to \Cref{subsec:sup_add_on} in the supplementary material.
{\cgf We rewrote our chair program using the \ourmethod{} add-on and found it to massively improve the manageability of crafting large programs compared to using only Blender's vanilla nodes. Quantitatively, the program written without the add-on had 819 nodes and 1K links while the program written with the add-on had 339 nodes and 462 links. Both variants are available in the supplementary material.}
Our add-on includes nodes that are grouped into categories: curves - open curves of various types; profile curves - planar closed curves that are extruded along an open curve; styles - nodes that determine the scale of the profile curve on each point along the open curve; meshes - usually a pre-prepared triplet (see~\Cref{fig:triplets}); sampling - nodes that sample a given curve at single or multiple intervals; symmetries - such as mirror or radial duplication; visualization - debug nodes, for example to visualize vectors or points.
Combined into a single framework, we show that these nodes allowed non-experts to craft complex procedural programs with only little guidance and in two different shape domains.

\subsection{\ourmethod{} user study} \label{subsec:user_study}

\begin{table}[t]
\centering
\small
\begin{tabular}{ l c c c c } 
\toprule
Shape Domain & Before & After & t-statistic & p-value \\
\midrule
Vase & 2.0 & \textbf{4.7} & 6.74 & 3.2e-5 \\
Ceiling lamp & 1.9 & \textbf{4.5} & 5.95 & 9.7e-5 \\
Novel domain & 1.6 & \textbf{3.75} & 5.82 & 1.2e-4 \\
\bottomrule
\end{tabular}
\caption{\textbf{Increased User Confidence in Building Procedural Programs.} We show the confidence of participants in our user study in building procedural models for a given shape domain before and after learning about \ourmethod{} through our user study. Confidence scores are given on a Likert scale (1-5, where 5 is the most confident) averaged over 12 participants. Users were more confident in their abilities to build \textit{novel} procedural programs even outside of the predefined domains.}
\vspace{-0.3cm}
\label{tab:confidence}
\end{table}

\begin{figure}[t!] %
    \centering
    \includegraphics[width=0.97\linewidth]{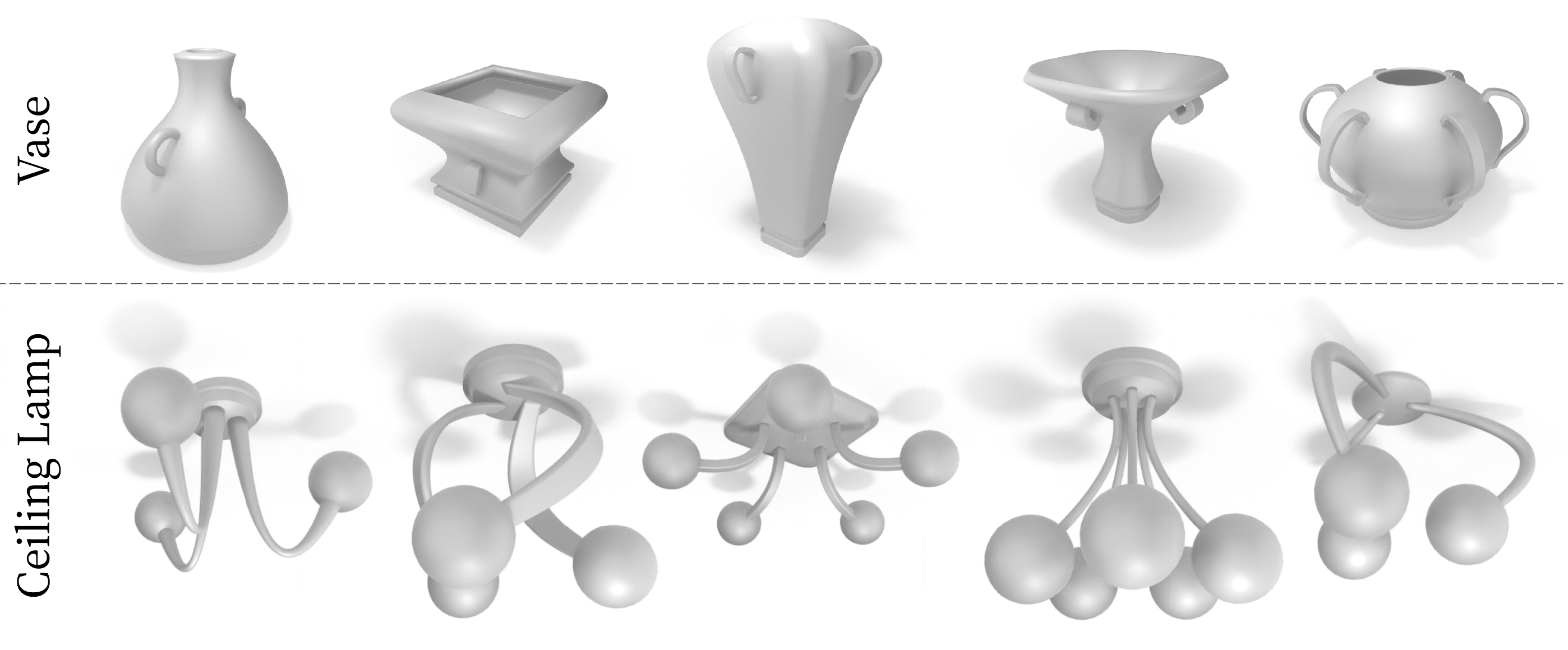}
    \caption{\textbf{User study shape programs.} Examples from the programs that our participants generated. The vase program has 20 input parameters and was generated as a step by step tutorial which also teaches the concepts of \ourmethod{}. The ceiling lamp program has 15 input parameters and was given as an exercise to craft the program without knowing the connections between the nodes. Both programs encode high variability and output high-quality meshes.
    }
    \label{fig:user_study}
\end{figure}

We conducted a user study to evaluate the benefits of using our procedural modeling methodology as captured by our \ourmethod{}.
The user study is comprised of two parts. In the first part, users learn about \ourmethod{}'s procedural modeling methodology and then follow a detailed guide to build a procedural program for vases (examples of shapes generated by the vase program are shown in the top row of \Cref{fig:user_study}). The vase program has over 20 input parameters, yet it is comprised of only 10 \ourmethod{} custom nodes and 5 built-in nodes.
In the second part of our user study, the participants are asked to create a program that generates \textbf{ceiling lamps}. During the ceiling lamp exercise, we supplied the participants with a high-level design of the ceiling lamp program that explains its requirements. The required nodes are hinted but with no direct context and no node links provided, the participant never saw the node graph in advance. The participants are asked to implement what they learned during the guided vase program and put together a ceiling lamp procedural program based on the requirements. The final ceiling lamp program has 15 input parameters and is comprised of 9 \ourmethod{} custom nodes and 5 built-in nodes. Questions were asked in four phases: before starting the user study (general knowledge), after completing the guided vase program, after completing the ceiling lamp exercise, and finally, after completing the user study.

\smallskip
\noindent \textbf{User study results.}
We had a total of 12 participants. Based on Likert-scale general knowledge questions, about half of our participants were familiar with Blender. However, none of them were familiar with Geometry Nodes. Eight of our participants predicted that building a vase program would take them more than a week. However, all our participants completed the guided vase program successfully and did so in an average time of 45 min. Similarly, eight participants predicted a week of work on a ceiling lamp program. However, our study shows that nine of our participants completed the exercise with no errors in an average of 32 min. Two other participants made a single mistake in the graph. The participants were able to implement {\cgf \ourmethod{} methodology and nodes to create an \textit{entirely} new domain}. This emphasizes the reusability of \ourmethod{} custom nodes across different shape domains.

To evaluate the effect of \ourmethod{} on non-expert users, we asked our participants about their confidence in building such programs before and after they completed the guide. We summarize the results in~\Cref{tab:confidence}. We observe a significant increase in the confidence of non-experts to craft new procedural programs.

\subsection{Shape Recovery} \label{subsec:exp_reconstruction}
We evaluate our method's ability to recover 3D shapes from samples in our test set. We also consider out-of-distribution examples, including shapes from %
COSEG~\cite{wang12aca}, real-world scan-based datasets such as ScanNet~\cite{dai2017scannet}, hand-drawn sketches, and sketches generated by CLIPasso~\cite{vinker2022clipasso} from images in the wild. Our supplementary contains additional experiments where we compare our work to \textcite{huang2017shapesynth}, show the importance of expressive shape parts, and compare the reconstruction accuracy between point cloud and sketch inputs.

{\cgf For quantitative evaluations, we use the bi-directional Chamfer Distance~\cite{chamfer}. The squared distance from each point in one point cloud to the closest point in the other point cloud. We use 10k randomly sampled points on the ground truth and reconstructed shapes.}

\begin{figure}[t!]
    \centering
    \includegraphics[width=0.92\linewidth]{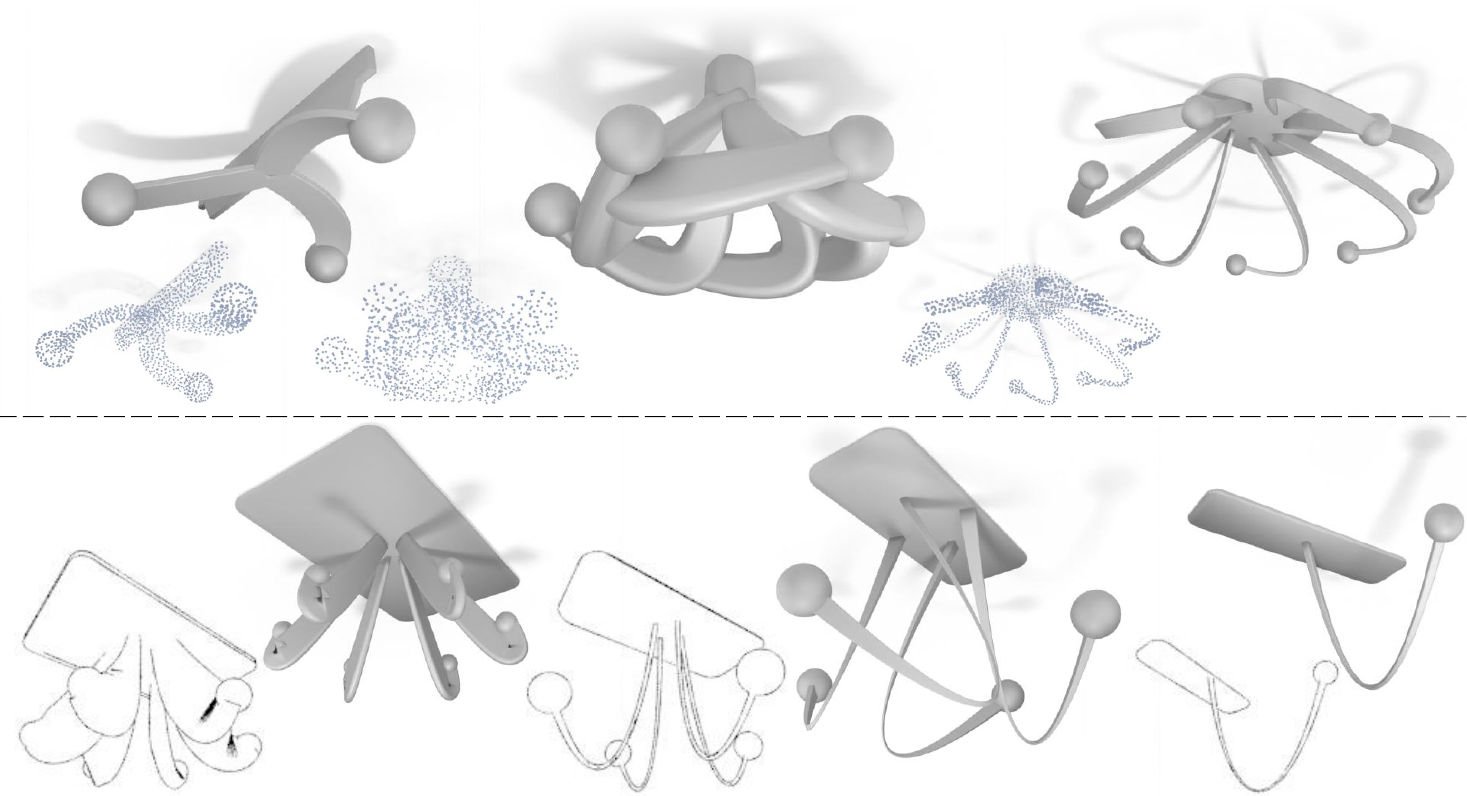}
    \caption{\textbf{Ceiling lamp on our test set.} \reva{Showing reconstructed ceiling lamps on the test set. We trained our model using the program that our non-expert participants made during the user study and we showcase the results of reconstruction from 3D point clouds and sketches.}}
    \label{fig:ceiling_lamp_test_set}
\end{figure}

\smallskip
\noindent \textbf{Results on our test set.}
We first verify qualitatively that our system can correctly reconstruct shapes given an input point cloud or sketch. In \Cref{fig:gallery}, we show examples of reconstructed shapes and their corresponding inputs. \reva{We do the same for the ceiling lamp program that was created by our user study participants in \Cref{fig:ceiling_lamp_test_set}. In both,} the reconstructions are visually similar to the input point clouds and sketches, confirming that our system can faithfully recover shapes for inputs within our data distribution, \reva{even for programs that were made by non-experts}.

\smallskip
\noindent \textbf{Comparison to \textcite{huang2017shapesynth}.}
{\cgf We test our suggested network against \cite{huang2017shapesynth}, which takes a different approach to predicting parameters that are the input to a procedural program. In their work, they train two different networks, one for predicting continuous parameters and another dedicated to predicting discrete parameters.
We trained \cite{huang2017shapesynth} on our training dataset until convergence. We implemented the loss function as depicted in their paper, and verified our actions with the authors. Then, we evaluated both methods on sketches from our test set.}
{\cgf \Cref{tab:huang_comparison} shows the average Chamfer Distances on all our domains separated by whether the sketches' camera angles were trained on or not. Please refer to a qualitative comparison in \Cref{subsec:sup_exp_shape_recovery}}.

{\cgf In \Cref{subsec:sup_failure_cases} we discuss a phenomenon where the shapes recovered from the novel camera angle are typically more rounded and in this experiment, we found that the work \cite{huang2017shapesynth} is susceptible to the same issue. However, while \ourmethod{} only suffers from that issue for the novel angle, the method described in \cite{huang2017shapesynth} seems to show the same issue even for trained camera angles. This explains the results shown in \Cref{tab:huang_comparison}.}

\begin{table}[t!]
\centering
\small
\setlength{\tabcolsep}{2pt}
\begin{tabular}{@{ } l l c c @{ }}
\toprule
Dataset & Method & Trained Angles $\downarrow$ & Novel Angle $\downarrow$ \\
\midrule
\multirow{2}{2em}{Chair} & \citeauthor{huang2017shapesynth} & 2.142e-3 & 8.030e-3 \\
& \ourmethod{} & 1.057e-3 & 8.075e-3 \\
\midrule
\multirow{2}{2em}{Vase} & \citeauthor{huang2017shapesynth} & 4.328e-3 & 9.271e-3 \\
& \ourmethod{} & 4.875e-3 & 9.788e-3 \\
\midrule
\multirow{2}{2em}{Table} & \citeauthor{huang2017shapesynth} & 4.051e-3 & 9.667e-3 \\
& \ourmethod{} & 2.497e-3 & 8.533e-3 \\
\midrule
\multirow{2}{2em}{\reva{Ceiling lamp}} & \reva{\citeauthor{huang2017shapesynth}} &  \reva{1.361e-2} & \reva{2.236e-2} \\
& \reva{\ourmethod{}} & \reva{1.403e-2} & \reva{2.272e-2} \\
\bottomrule
\end{tabular}
\caption{\textbf{Comparison to \citeauthor{huang2017shapesynth}~\citeyearbrackets{huang2017shapesynth}.} Comparing the average Chamfer Distance on our test set to \citeauthor{huang2017shapesynth}~\citeyearbrackets{huang2017shapesynth}. We explain the result by the fact that our method is more resilient to shapes that are not rotationally symmetric, thus, the Chamfer Distance for both the chair and table domains is improved by \ourmethod{}.  In both cases, we observe a highly statistically significant result (with a P-value of 2.2e-10 and 1.8e-6 respectively). Additionally, both methods perform similarly when the sketches are drawn from a novel angle.
}
\vspace{-0.2cm}
\label{tab:huang_comparison}
\end{table}

\begin{figure}[t!] %
    \centering
    \includegraphics[width=0.99\linewidth]{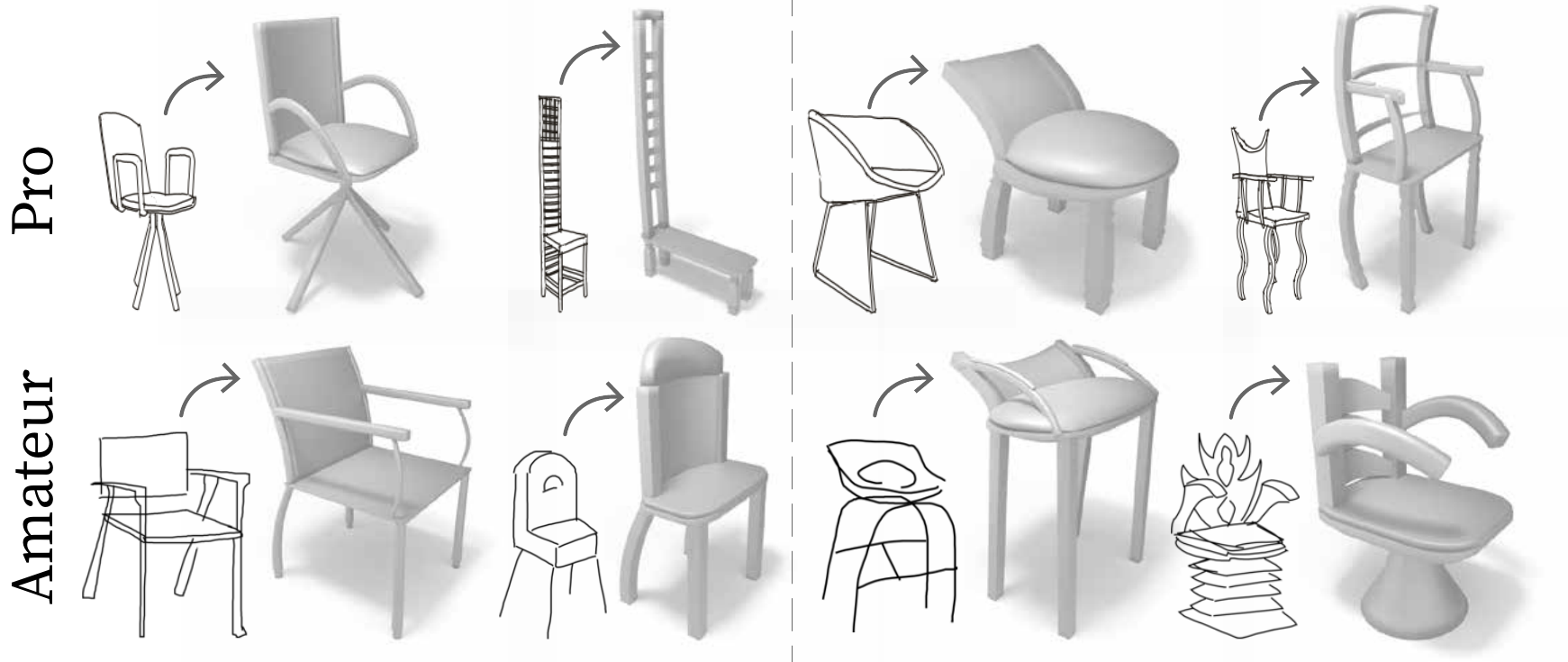}
    \caption{\new{\textbf{Reconstruction from amateur and professional sketch datasets.} We use sketches from AmateurSketch and ProSketch chairs datasets \cite{sketchx} and show qualitative results for 3D shape recovery using \ourmethod{}. On the left, we show reconstruction for samples that are within the distribution of shapes the chair program can generate; on the right, we show recovery for shapes that are out of distribution. \ourmethod{} does incur errors such as the dimensions of the second professional sketch or the handles in the third amateur sketch. However, even when the samples are particularly challenging, \ourmethod{} is able to capture the main geometrical attributes in the recovered shape.
    }}
    \label{fig:sketchx}
\end{figure}

\smallskip
\noindent \textbf{Reconstruction from out-of-distribution sketches.}
We conduct a qualitative evaluation of sketch inputs from AmateurSketch-3DChair and ProSketch-3DChair datasets, both datasets contain sketches of shapes in ShapeNet~\cite{shapenet2015} drawn by artists from various angles as part of the work Sketch-Based 3D Shape Generation~\cite{sketchx}. AmateurSketch-3DChair is a dataset created by amateur artists and the ProSketch-3DChair is created by professionals. \Cref{fig:sketchx} shows that while some discrepancies between the sketches and their reconstructions exist, overall \ourmethod{} is able to capture key attributes from both sketch types even for shapes that are considered out-of-distribution. We report an average Chamfer Distance of 0.042 for the AmateurSketch dataset (1005 shapes) and 0.049 for the ProSketch dataset (500 shapes).

\begin{figure}[t!] %
    \centering
    \includegraphics[width=0.95\linewidth]{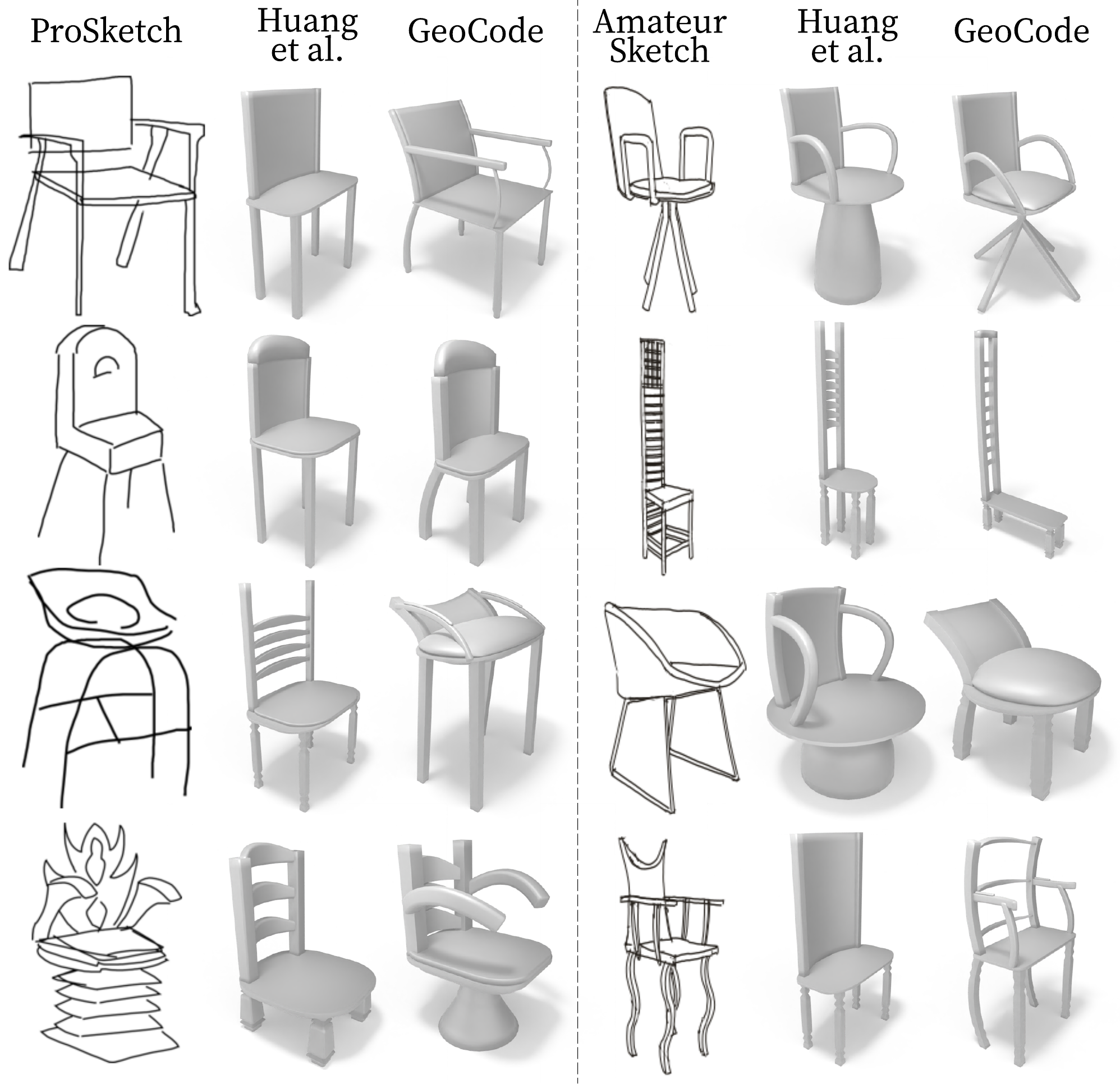}
    \caption{\textbf{Qualitative comparison to \textcite{huang2017shapesynth} {\sig on out-of-distribution sketches.}} We use AmateurSketch and ProSketch sketches to show a qualitative analysis of 3D shape recovery between the two methods. The top two rows are samples that we consider in-distribution, while the bottom two rows are samples that are out-of-distribution. We find that \ourmethod{} is more capable of capturing important geometric details.}
    \label{fig:out_of_distribution_comparison}
\end{figure}

\noindent \textbf{Qualitative comparison to \textcite{huang2017shapesynth} on out-of-distribution sketches.}
We go on to compare \ourmethod{} to \textcite{huang2017shapesynth} on reconstruction from out-of-distribution sketches. \Cref{fig:out_of_distribution_comparison} shows a qualitative comparison of the same sketches used in \Cref{fig:sketchx} while comparing both methods. Both methods experience difficulties, however, the reconstruction results of \ourmethod{} far exceed those of \textcite{huang2017shapesynth}. We direct the readers' attention to the bottom two rows where the samples are considered out-of-distribution since the programs cannot generate a shape that closely matches those samples. In these two rows, the advantage of \ourmethod{} is even more evident.

\begin{figure}[t!]
    \centering
    \newcommand{\blue}{\color[rgb]{0.557,0.851,0.937}}
    \newcommand{\purple}{\color[rgb]{0.898,0.549,0.918}}
    \newcommand{\green}{\color[rgb]{0.616,0.91,0.545}}
    \includegraphics[width=0.95\linewidth]{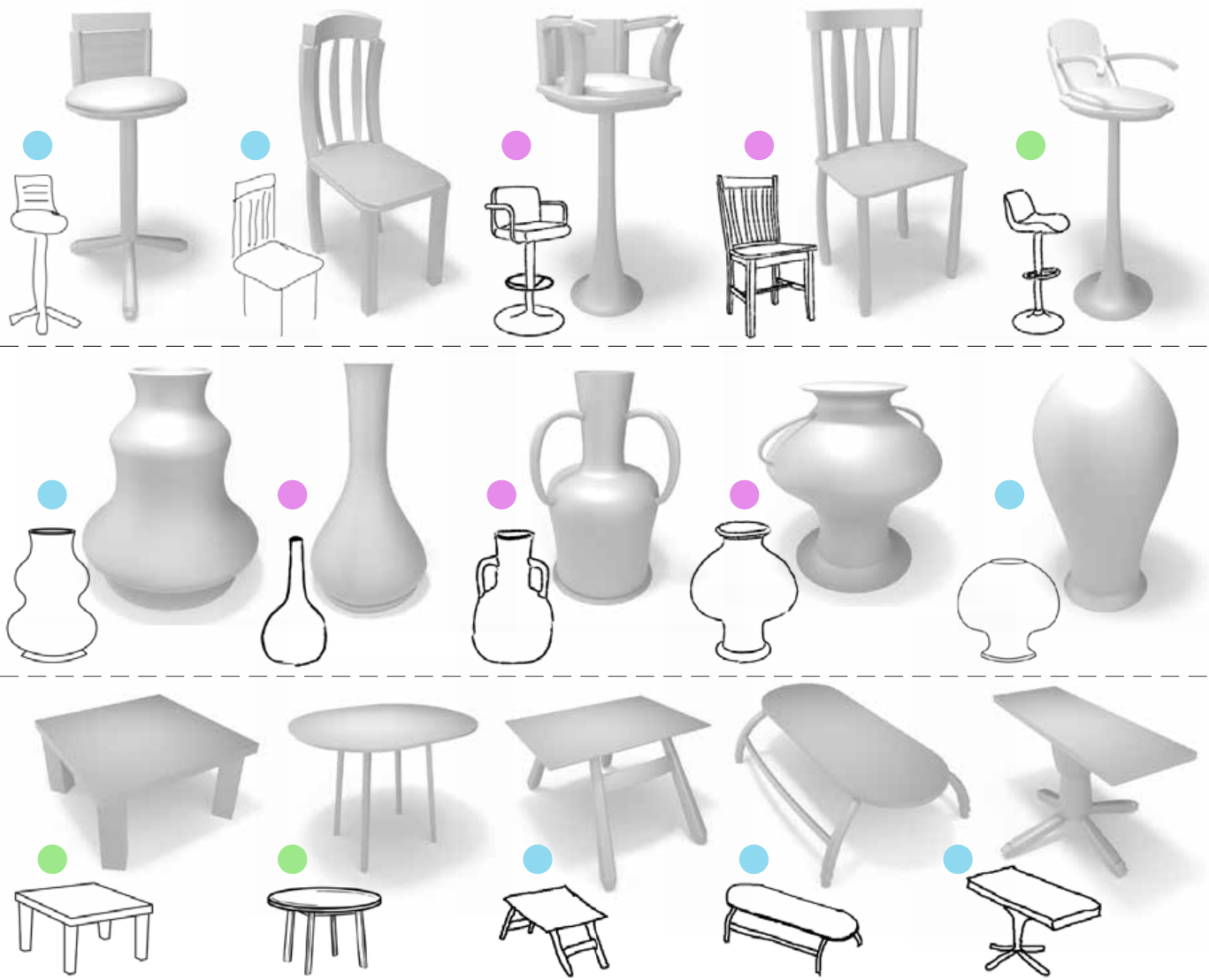}
    \caption{\textbf{Hand-drawn sketches.} We show the ability of \ourmethod{} to recover an editable 3D shape from {\blue\textbf {free-form sketches}}, {\purple\textbf {sketch outlining}}, and  {\green\textbf {sketches in the wild}}. Our system produces a 3D shape that captures the main attributes of the input sketches.}
    \label{fig:traced}
\end{figure}

\noindent \textbf{Reconstruction from hand-drawn sketches.}
~\Cref{fig:traced} shows reconstruction examples from sketches that we drew by hand (free-form sketches), pictures from the web that we converted into sketches by outlining the shapes (sketch outlining), and sketches from the web (sketches in the wild). Our system is able to generalize and capture the main features from the sketches and produce shapes that are visually similar to the input sketches. These sketches exhibit different styles and are drawn from different angles compared to the sketches we trained on.

\begin{figure}[t!]
    \centering
    \includegraphics[width=0.95\linewidth]{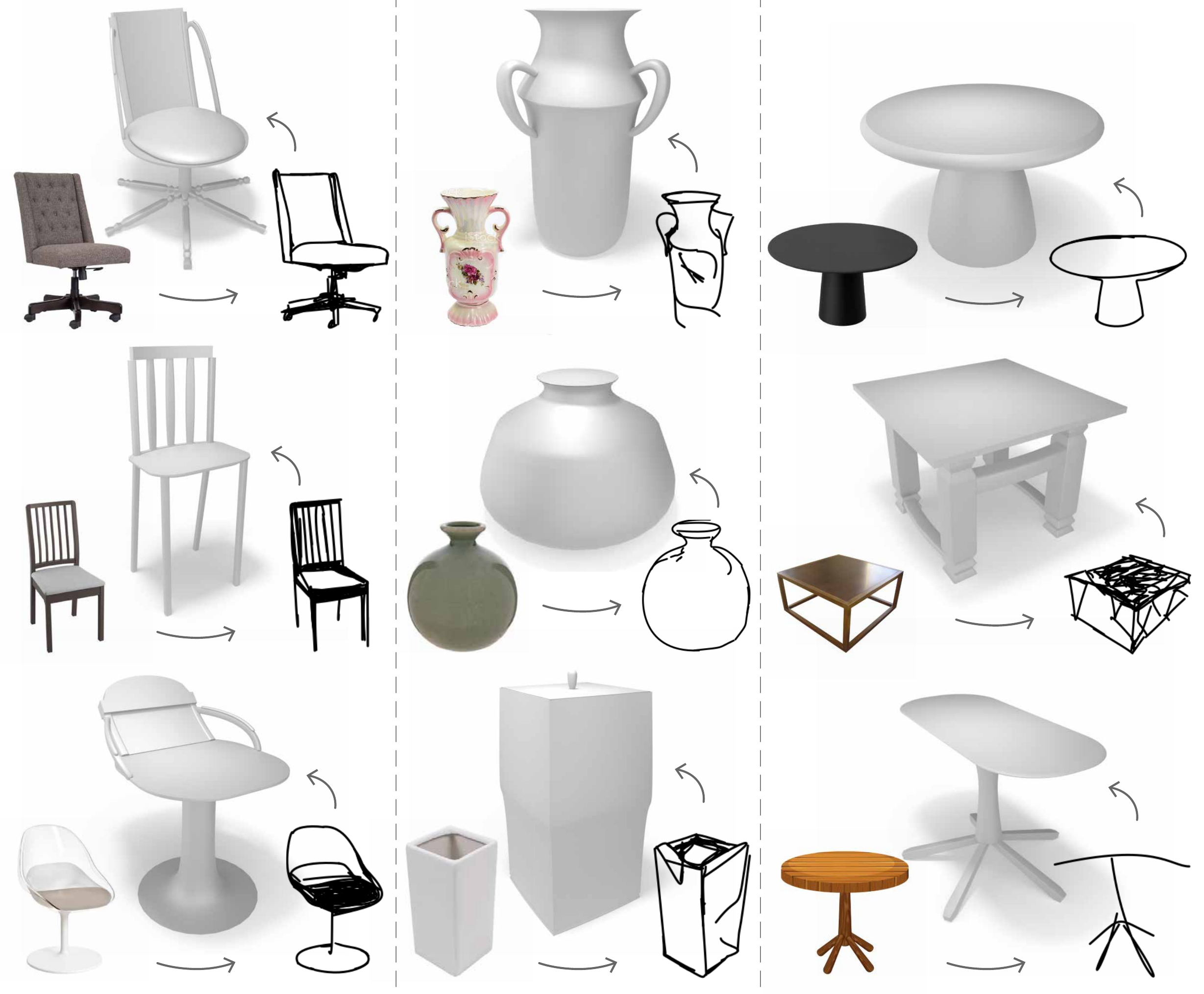}
    \caption{\textbf{Sketches from images.} Starting with an input image, we automatically create a sketch using CLIPasso~\cite{vinker2022clipasso}. Our method generates editable shapes that closely match the original images and the corresponding sketches that have a variety of styles.}
    \label{fig:clipasso}
\end{figure}

\noindent \textbf{Reconstruction from CLIPasso sketches.}
~\Cref{fig:clipasso} shows reconstruction examples from sketches generated by CLIPasso~\cite{vinker2022clipasso}. We use CLIPasso to convert images of chairs, vases, and tables found in the wild to sketches with various numbers of strokes. The sketches produced by CLIPasso are noisy and often contain artifacts, these make them visually different from sketches found in our datasets. %
We observe that, overall, the shapes reconstructed from our system are sensible and correctly recover features from the shapes in the original images. This suggests that our network is able to express shapes with human-interpretable parameters given an out-of-distribution input.

\smallskip
\noindent \textbf{Reconstruction from out-of-distribution point clouds.}
To further demonstrate the generalization capability of \ourmethod, we choose three known 3D shape datasets and reconstruct shapes from these datasets for point cloud inputs.
First, we consider two real-world scan-based datasets, ScanObjectNN~\cite{uy-scanobjectnn-iccv19} and ScanNet~\cite{dai2017scannet}, which include LiDAR scans of objects and scenes, respectively. For ScanObjectNN, we use the point clouds from the main dataset. For ScanNet, we extract objects from the scene using the provided segmentation masks and randomly sample point clouds from them. We also consider ShapeNet~\cite{shapenet2015} and produce point clouds by randomly sampling its shapes. For all the datasets, we end up with point clouds of 2048 points.

\begin{figure}[t!]
    \centering
    \includegraphics[width=0.99\linewidth]{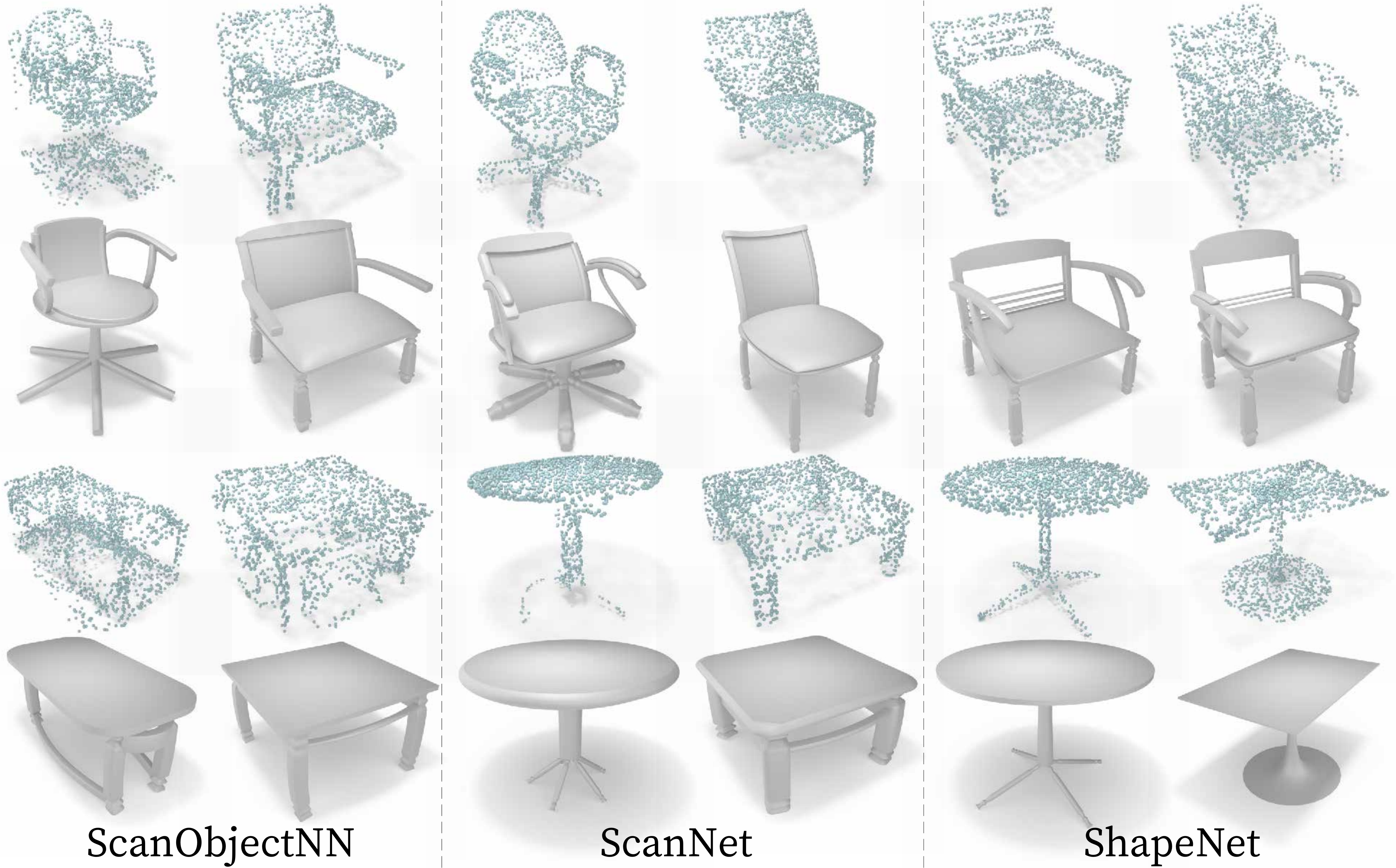}
    \caption{\textbf{Generalization to out-of-distribution point clouds.} We show reconstruction results of \ourmethod{} for point clouds from various datasets. We use scanned objects from ScanObjectNN~\cite{uy-scanobjectnn-iccv19} and ScanNet~\cite{dai2017scannet} and also test on point clouds from ShapeNet~\cite{shapenet2015}. Our method outputs plausible reconstructions, despite hindered point cloud quality that is characterized by noise and missing parts.}
    \label{fig:various_pc_reconstruction}
\end{figure}

In~\Cref{fig:various_pc_reconstruction} we show the reconstruction for two domains, chairs, and tables. Our method is able to recover shapes that capture the main geometric features of the out-of-distribution samples. Moreover, point clouds produced from scanned objects are contaminated by noise and partiality. Still, \ourmethod{} is able to produce compelling results in these cases as well.

\subsection{Shape Editing} \label{subsec:exp_edits}
Beyond shape recovery, we show that our interpretable parameter space allows for easy shape editing. This property is exemplified by shape mixing and shape interpolation.

\begin{figure}
    \centering
    \newcommand{\orange}{\color[HTML]{ff9644}}
    \newcommand{\blue}{\color[HTML]{5eb1ff}}
    \includegraphics[width=0.99\linewidth]{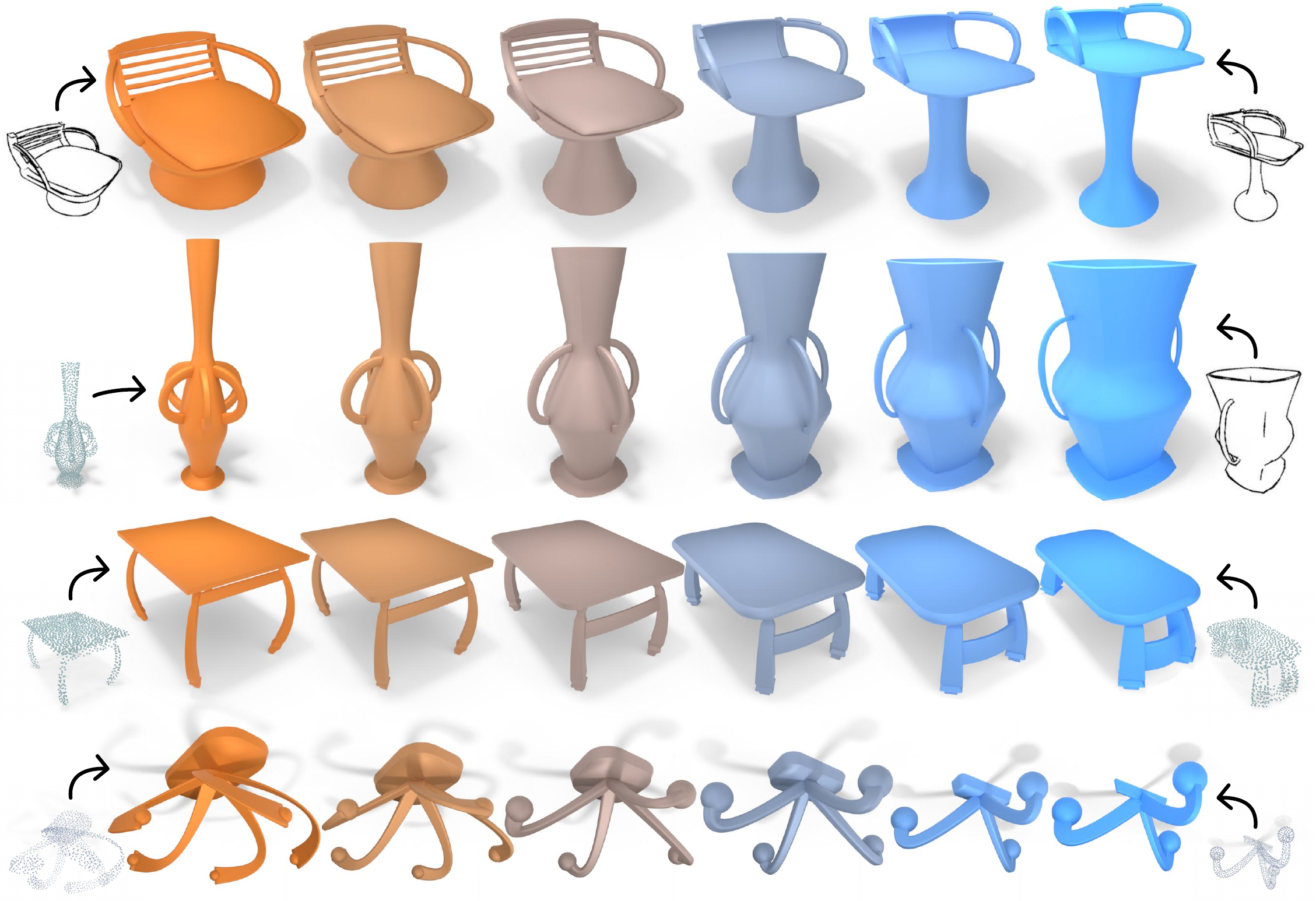}
    \caption{\textbf{Shape interpolation.} We uniformly interpolate between two ({\orange\textbf{left}} and {\blue\textbf{right}}) \ourmethod{} shapes. Interpolations contain gradual changes (continuous parameters) as well as sharp structural changes (discrete or binary parameters).}
    \label{fig:interpolate}
\end{figure}

\begin{figure}
    \newcommand{\source}{\color[rgb]{1,0.64,0}}
    \newcommand{\features}{\color[HTML]{5eb1ff}}
    \newcommand{\second}{\color[HTML]{959595}}
    \centering
    \includegraphics[width=0.90\linewidth]{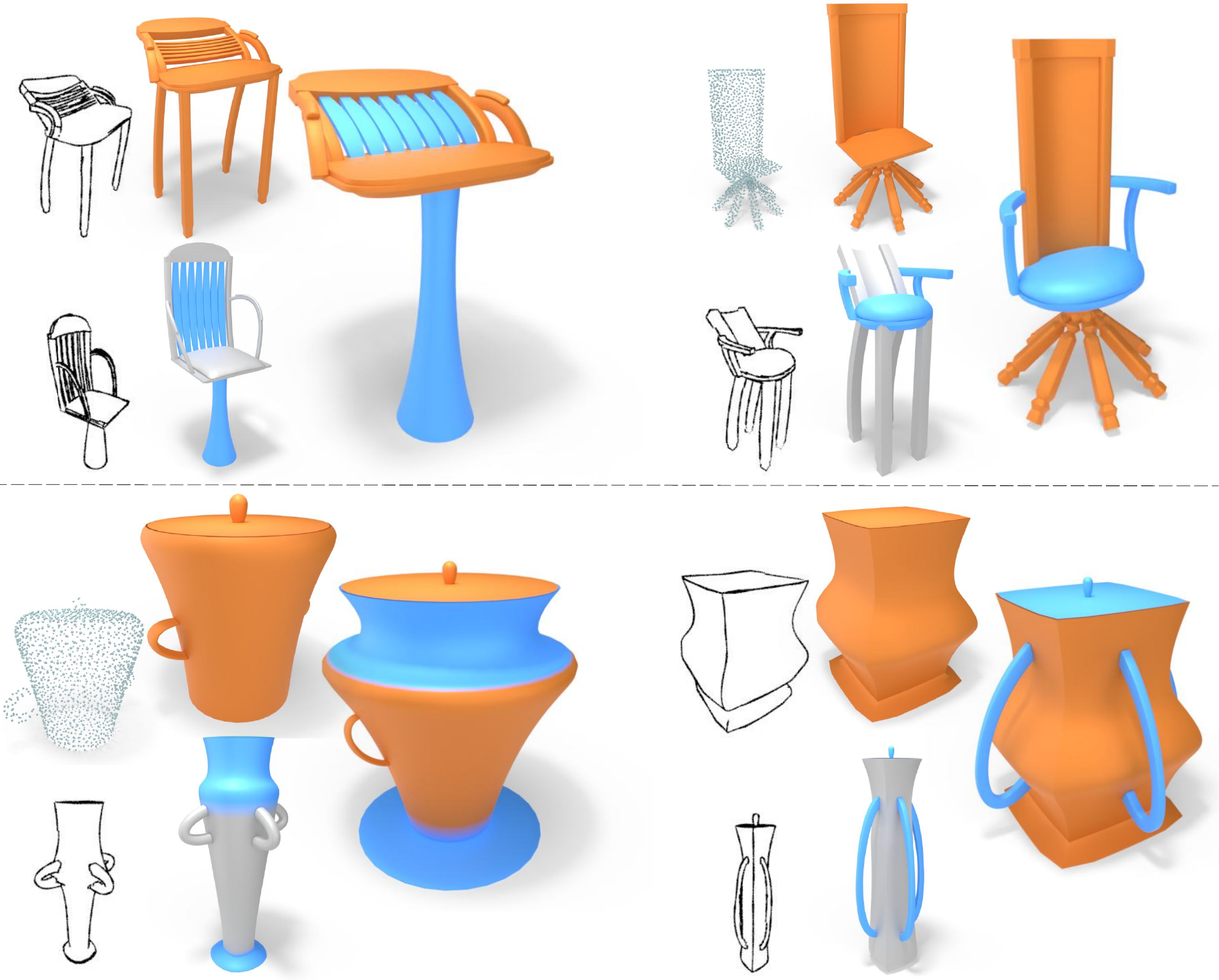}
    \caption{\textbf{Shape mixing.} We demonstrate shape mixing by taking {\features \textbf{geometric features}} from a {\second \textbf{source shape}} and copying them to a {\source \textbf{target shape}}. \ourmethod{} will either add the selected geometric features to the target shape (\eg, handles are added to the second vase), or it will override them in the target shape (\eg, cross rails in the backrest of the first chair are replaced with vertical rails). In both cases, the structural integrity of the shape is maintained.}
    \label{fig:shape_mixing}
\end{figure}

\smallskip
\noindent \textbf{Shape mixing.}
We show that our system can mix shapes by selecting a set of parameters from the human-interpretable parameter space representation of a source shape and copying them to a target shape.
~\Cref{fig:shape_mixing} shows mixing examples between pairs of shapes reconstructed from point clouds and sketches. The resulting shapes are based on the first shape with the addition of selected parts from a second shape. The final mixed shape is structurally valid and physically plausible.

\smallskip
\noindent \textbf{Shape Interpolation.}
\ourmethod{} is effective in interpolating between shapes.~\Cref{fig:interpolate} shows interpolation between pairs of shapes reconstructed from point clouds and sketches where we interpolate across all the parameters for $\alpha \in [0,1]$ in 0.2 increments. In our supplementary, we demonstrate interpolation over selected set geometric features.

\subsection{Ablation Study} \label{subsec:exp_ablation}

{\cgf We perform an ablation study to support our choice of using classification for float values. Please refer to \Cref{subsec:sup_exp_ablation} for an additional ablation study.}

\smallskip
\noindent \textbf{Classification \vs regression.} {\cgf In our work, we supported discrete, Boolean, and continuous parameters. We explained in \Cref{sec:method} that continuous parameters are discretized and that the loss function is cross-entropy. In this ablation experiment, we test the reconstruction in terms of average Chamfer Distance while comparing the use of classification \vs regression for the continuous parameters.}
\par
\begin{figure}
    \centering
    \includegraphics[width=0.99\linewidth]{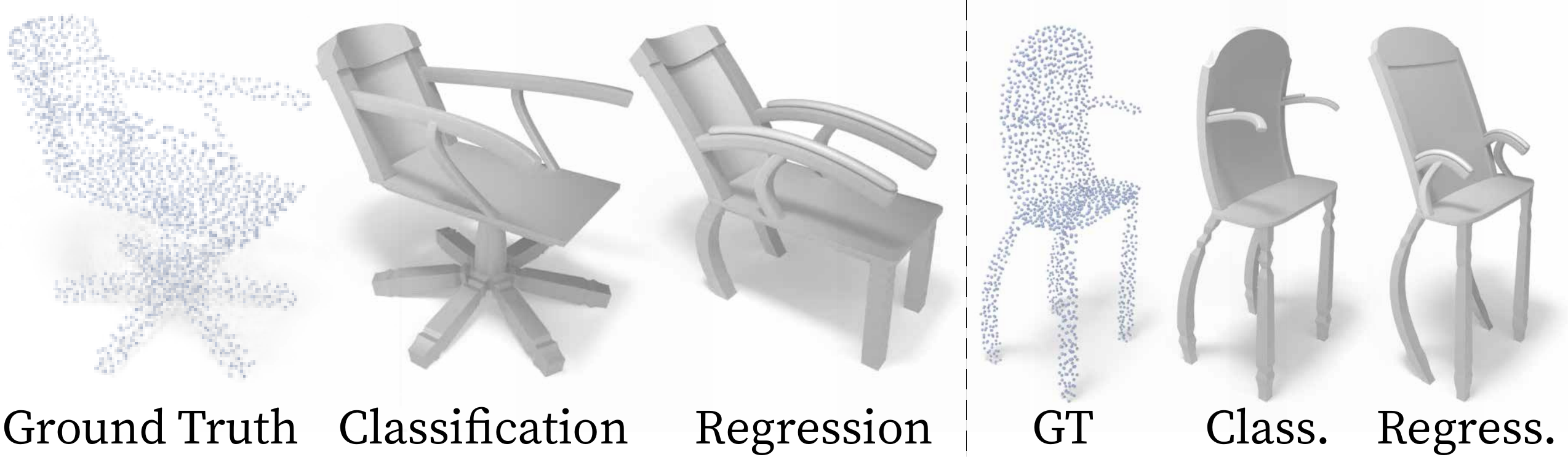}
    \caption{\new{\textbf{Classification \vs regression.} {\cgf Qualitative comparison between reconstruction when continuous parameters are predicted using regression \vs discretizing them and predicting them using classification. We show the improved reconstruction on selected chair samples favoring prediction using classification.}}}
    \label{fig:failure_cases_regression}
\end{figure}

{\cgf In \Cref{tab:regression} we can see that using classification for all the parameters yields better average Chamfer Distance results. In \Cref{fig:failure_cases_regression} we show two examples from the chair domain that show the difference between the two methods. Another important part of this experiment is explained in \Cref{subsec:sup_dataset} where we discuss the \textit{part existence label} and its role in this experiment.}

\begin{table}
\centering
\begin{tabular}{l c c}
\toprule
Method & Point Cloud $\downarrow$ & Sketch $\downarrow$ \\
\midrule
Classification & \textbf{0.00040} & \textbf{0.00106} \\
Regression & 0.00198 & 0.00364 \\
\bottomrule
\end{tabular}
\caption{\textbf{Classification \vs regression.} {\cgf We train the chair network in two different ways. Classification: continuous parameters are discretized and the loss function is cross-entropy, as described in \Cref{sec:method}. Regression: we employ cross-entropy loss for discrete and binary parameters, but we use regression to predict continuous ones. We can see that the Chamfer Distance is better overall when using classification loss for all the parameters. The results are highly statistically significant, with P-values of 1.9e-24 and 7.2e-50 for point cloud and sketch reconstruction respectively.}
}
\vspace{-0.3cm}
\label{tab:regression}
\end{table}

\reva{\subsection{Case study with an expert}} \label{subsec:exp_ablation}
\reva{We discussed our work with a computer graphics artist with over 10 years of experience in the industry and approximately one year of experience with Blender Geometry Nodes. Firstly, the expert confirmed that, to their knowledge, they are 'not aware of any other framework that resembles what GeoCode add-on does'.}

\reva{The expert correctly noted that 'Non-experts will need at the very least a short introduction to Blender and Geometry Nodes before jumping into using GeoCode add-on'. This is the very reason that our user study (refer to \cref{subsec:user_study}) required an introduction to Blender and our add-on to provide basic tools to our participants. They proceeded to say, 'Experts would probably find it extremely logical and intuitive. GeoCode itself is quite easy to use for a person who knows their way around the Geometry Nodes system'.}

\reva{When asked about part-based shapes, and taking the Cabinet example (see \Cref{fig:structural_integrity_discussion}), the expert said 'Given the task of creating such a complex program, I would probably find it unmanageable without using a framework such as GeoCode'.}

\reva{We also asked about what improvements they thought could be made to GeoCode within the confines of Blender. The answers revolved around pain points that many CG artists deal with on a daily basis - optimizing the topology for game assets, UV unwrapping, and selectable preset materials. All of which can form future improvements to our work.}

\reva{Finally, the expert stated that 'Experts can even use GeoCode add-on and programs as a resource for learning Geometry Nodes'.}

\reva{\section{Limitations}} \label{sec:limitations}

\reva{Building the chair, vase, and table programs which have 59, 39, and 36 parameters took roughly 2 weeks, a single week, and 5 days, respectively. While this is a significant amount of time, these programs were crafted with the concepts of \ourmethod in mind, but before we created the Blender add-on. With our Blender add-on, we were able to re-program the chair program in a mere 4 days. The ceiling lamps program (19 parameters) from our user study was built by non-expert participants in an average time of 32 min after approximately 40 min of a guided tutorial. In addition, to further strengthen the trust in \ourmethod's ability to extend to new domains we created the cabinet program (25 parameters) in a record time of just 2 days.}

\reva{\ourmethod{} does not guarantee the structural integrity of the shapes simply by using the add-on and the methodology that we presented in this work. However, a good design will lead to valid shapes even when introduced with mistakes, we examine this using our cabinet program when discussing the structural integrity in \ref{subsec:procedual_modeling_methodology}. In addition to providing useful nodes in their own right, we also handle many pitfalls that programmers may encounter when using vanilla Geometry Nodes, such as unexpected curve direction flips, alignment with axes, attachment to rotated elements, and much more.}

\reva{We also recognize that \ourmethod{} will not be useful for every conceivable shape, for example, we would not recommend it for generating organic shapes such as faces or some animals. It is intended to be used for part-based shapes.}

\reva{Finally, while our network's predictions from point clouds are impressive, our predictions from sketches are only improving upon the existing methods in certain conditions, and interestingly, the predictions from the novel angle suffer from the tendency for rotational symmetry as we explained in \cref{subsec:sup_failure_cases}.}

\section{Conclusion}

In this paper, we presented \ourmethod{}, a novel method that aims to lower the barrier for crafting shape programs that can generate 3D shapes using a human-interpretable parameter space. 
We show the effectiveness of shape programs by building three domain-specific procedural programs controlled by intuitive parameter spaces and training a neural network to predict the parameter representation for an input point cloud or sketch. We showed that our system produces structurally valid 3D geometry and enables editing of the resulting shape easily and intuitively. 
Our user study demonstrated that our framework enables users with no prior 3D modeling experience to produce shape programs. Specifically, users can take the core components from these programs and build a completely novel class (ceiling lamps), which we did not create a program for.

In the future, we are interested in incorporating additional interactive features, such as viewport manipulations and edits to shapes. Incorporation of additional edit modalities (such as drawing a part on a screen) combined with our visual programs may provide additional opportunities for program exploration and design.



\section{Acknowledgments}

We thank the University of Chicago for providing the AI cluster resources, services, and the professional support of the technical staff. This work was also supported in part by gifts from Adobe Research. Finally, we would like to thank R. Kenny Jones, Chen Dudai, Noam Sahar, and the members of 3DL for their thorough and insightful feedback on our work.

\printbibliography



\clearpage

\appendix

\twocolumn[{%
 \centering
 {\textbf{\Large{Supplementary Material for GeoCode: Interpretable Shape Programs}}}
 \vspace{2em}
}]

We provide additional information about \ourmethod{}. \Cref{sec:sup_additional_resources} describes our Blender add-on that the participants used in our user study and three complementary videos. In \Cref{sec:sup_experiments}, we provide additional experiments, evaluate the robustness of \ourmethod{}, present an ablation study, and discuss failure cases. Lastly, \Cref{sec:sup_impl} elaborates on our implementation details, including our dataset creation, network architecture, and training scheme. 

\section{Additional Resources} \label{sec:sup_additional_resources}

\subsection{Blender \ourmethod{} Add-On} \label{subsec:sup_add_on}

\smallskip
An inseparable part of our contribution is the Blender \cite{blender} add-on, which we validated with the user study to allow non-experts to generate versatile shape programs for a novel domain of ceiling lamps. Please refer to \Cref{subsec:user_study} for more information. In this section, we detail the nodes that we implemented in the design philosophy of \ourmethod{}.

\noindent Staying true to \Cref{fig:triplets} we provide several custom nodes for the \textit{Curve}, each with different behavior and controls:
\begin{itemize}[topsep=3pt,itemsep=1pt,partopsep=1pt, parsep=1pt]
    \item CurveHandle - handle with two attachment points to another given curve, this is a complex curve that maintains structural validity by avoiding collisions with the given curve.
    \item CurveQuadraticPolar
    \item CurveQuadraticCartesian
    \item CurveFixedTop
    \item CurveFixedMiddle
    \item CurveFixedBottom
    \item CurveBezierSegment
    \item CurveArc
\end{itemize}

\noindent We found that a single \textit{Profile Curve} can answer all our needs:
\begin{itemize}[topsep=3pt,itemsep=1pt,partopsep=1pt, parsep=1pt]
    \item ProfileCurveRectangleFillet
\end{itemize}

\noindent We provide several custom nodes for \textit{Style}:
\begin{itemize}[topsep=3pt,itemsep=1pt,partopsep=1pt,parsep=1pt]
    \item StyleSolidify - contrary to all other style nodes, this is the only node that is not receiving a \textit{Curve} and a \textit{Profile Curve}. Instead, it receives a mesh and applies thickness to it.
    \item StyleLeg - new styles are easy to add since we make use of a curve editor node within Blender, which is easy to adjust. This is true for other nodes as well.
    \item StyleSeatFrame
    \item StyleRail
    \item StyleMiddleExpandContract
    \item StyleCushion
    \item StyleBezierSegment
    \item StyleAlignedBevelXY
\end{itemize}

\noindent Sampling curves to be used as attachment points is an important design methodology of \ourmethod{}. We provide the following \textit{Sample} nodes:
\begin{itemize}[topsep=3pt,itemsep=1pt,partopsep=1pt, parsep=1pt]
    \item SampleCurve - given a percent value, return a point at the respective length along the curve.
    \item SampleCurveArray - given a requested number of points, return points at equal distances along the curve (with an option to set margins).
    \item SamplePoints - given a set of points and an index, return a single point that corresponds to the provided index.
\end{itemize}

\noindent \textit{Symmetry} nodes are intuitive and were used as rules in many previous works, they are also an integral part of our design:
\begin{itemize}[topsep=3pt,itemsep=1pt,partopsep=1pt, parsep=1pt]
    \item SymmetryRadial - the number of duplicates and the radius are given as input.
    \item SymmetryMirror - the axis is chosen as input.
\end{itemize}

\noindent To further assist novice users, and to provide examples for using the previous nodes with \ourmethod{} design philosophy, we provide several \textit{Mesh} nodes that are in essence, pre-prepared triplets (refer to \Cref{fig:triplets}):
\begin{itemize}[topsep=3pt,itemsep=1pt,partopsep=1pt, parsep=1pt]
    \item MeshTopRail
    \item MeshStraightFrame
    \item MeshSeat
    \item MeshQuadraticPolarFrame
    \item MeshQuadraticCartesianAlignedXYFrame
    \item MeshQuadraticCartesian
    \item MeshLeg
    \item MeshFrameFixedBottom
    \item MeshCushion
    \item MeshCrossRail
\end{itemize}

\noindent Programming required debugging and visual programming with node-based systems, are no exception in that regard. As our programs become more complex, there is a need to visualize our work, we provide some basic custom nodes that we felt were missing from Blender's Geometry Nodes Editor \cite{blender-geometry-nodes}:
\begin{itemize}[topsep=3pt,itemsep=1pt,partopsep=1pt, parsep=1pt]
    \item VisualizeVectorsAtPoints
    \item VisualizeVector
\end{itemize}

\noindent Finally, some Util nodes that repeated several other custom nodes are also provided. Those were mostly used to align the curve direction to be intuitive as possible, as Blender usually picks the direction based on other attributes of the curve, and as the curve changes, its direction can suddenly flip, that is a behavior that we aimed to eliminate with \ourmethod{}:
\begin{itemize}[topsep=3pt,itemsep=1pt,partopsep=1pt, parsep=1pt]
    \item UtilDashed
    \item UtilAlignCurveToStartPoint
    \item UtilAlignCurveTiltToAxis
\end{itemize}

\subsection{Complementary Videos} \label{subsec:sup_complementary_videos}

\smallskip
\noindent \textbf{Demo video.} As part of the supplementary, we provide a video (\textbf{GeoCode Demo.mp4}) showing the variety of shapes that our programs can produce. In addition, the video will help the reader understand the ease of controllability that the method provides, as well as show how the structural integrity is maintained continuously while editing the shapes.
\par
For a short length at the beginning of the video, we also show the ease at which shapes can be changed using human-interpretable parameters, by showing the shape along with the actual values of selected parameters of the chair:

\begin{itemize}[topsep=4pt,itemsep=1pt,partopsep=1pt, parsep=1pt]
    \item Cross Rail Count (discrete) - controls the number of cross rails appearing on the backrest of the chair.
    \item Armrest (boolean) - toggles the existence of the armrests, when set to \textit{True} the armrests will appear while maintaining the structural integrity of the current chair shape.
    \item Armrest Height (continuous) - controls the attachment point of the base of the armrests to the backrest of the chair. Higher values, mean the armrests are connected to the frame at a higher position. The value is treated as a percentage of the height of the frame, for example, a value of 0.5 will place the armrests at an equal height between the seat and the top of the frame.
    \item Seat Roundness (continuous) - controls the shape of the seat. The higher this value is, the more rounded the seat shape will be. The video helps to show that the entire chair is adjusted automatically to fit the changing seat shape and maintain the structural integrity of the chair. We show how this is achieved when we describe our method in \Cref{fig:program_inside_look} of the main paper.
    \item Seat Height (continuous) - controls the height of the seat. The legs, the frame, and other components of the chair, like the cross rail, are all adjusted automatically to fit the seat position in a continuous manner.
    \item Backrest Curvature (continuous) - this parameter controls how rounded the cross rails and top rail are. A value of 0.0 means that the backrest is completely flat, while higher values will make the backrest appear more rounded. {\new This will affect the backrest across all its types, cross rails, vertical rails, and solid backrest while maintaining a physically valid shape.}
    \item Backrest Slant (continuous) - allows adjusting how slanted the backrest is. A value of 0.0 means the backrest is perpendicular to the ground (or equivalently, to the seat), while higher values make the backrest lean backward in a curved fashion.
\end{itemize}

\smallskip
\noindent \textbf{Simulation demonstration.} In \Cref{sec:sup_experiments}, we discuss our \textit{structural stability} metric which checks that a given shape is structurally valid by testing two things, checking for the existence of loose parts, and dropping the shape from a height onto a flat plane and testing that the height of the shape has not changed after the drop {\new by more than 2\%}. In the attached video (\textbf{GeoCode Simulation.mp4}) we show 15 samples tested in that simulation, from all three domains, chair, vase, and table. The drop height was increased for the purpose of visual demonstration, from the 5\% that we used when creating \Cref{tab:stability} to 20\% of the shape's longest dimension. {\new The third video (\textbf{GeoCode Simulation Unstable.mp4}) shows an example of chairs that failed in that physics simulation.}

\section{Additional Experiments} \label{sec:sup_experiments}

\subsection{Shape Recovery} \label{subsec:sup_exp_shape_recovery}

\subsection{Comparisons \& Evaluation} \label{subsec:sup_comparison_other_methods}

In this section, we evaluate our network's ability to recover 3D shapes from either a point cloud or a sketch and qualitatively compare our work to a similar work~\cite{huang2017shapesynth}. Additionally, we evaluate the expressiveness of our method compared to other methods that use coarse bounding boxes as shape parts~\cite{mo2019structurenet, shapeAssembly}. The task that these solve is harder as they infer the relations between the parts either as a graph or a written program. However, we use them to demonstrate the benefit of using more expressive shape parts that are based on curves and generally on our procedural modeling methodology~\Cref{subsec:procedual_modeling_methodology}.

\begin{figure}[t!]
    \centering
    \includegraphics[width=0.95\linewidth]{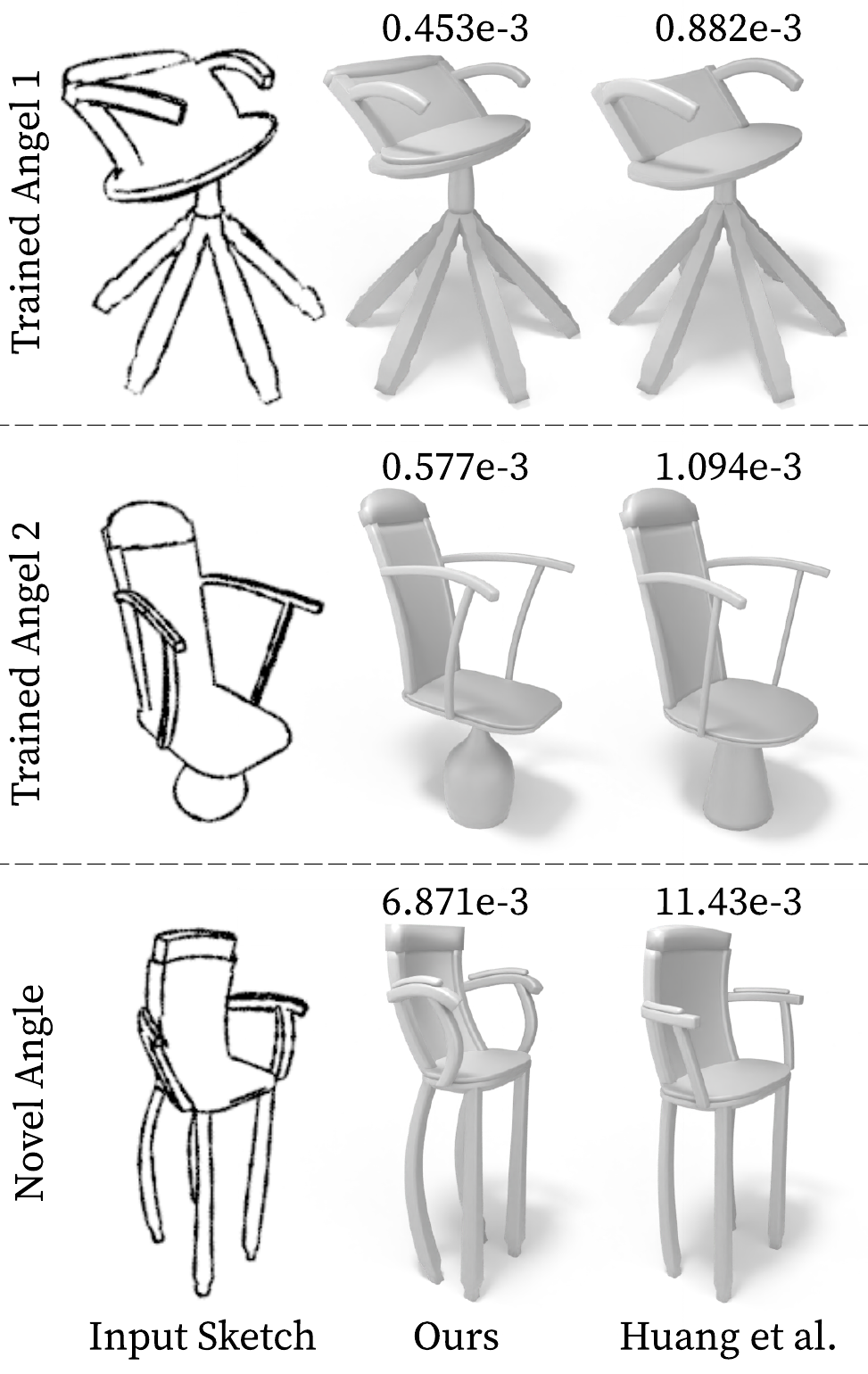}
    \caption{\textbf{Comparison to \cite{huang2017shapesynth}.} Qualitative comparison of \ourmethod{} and \cite{huang2017shapesynth} after they were both trained on our training set. \cite{huang2017shapesynth} does a great job at predicting the parameters, however, \ourmethod{} seems to perform better. In the first row, \ourmethod{} captures the top rail more accurately. In the second row, the seat roundness captured by \ourmethod{} is much more precise. Finally, in the third row, the bend in the legs predicted by \ourmethod{} better matches the ground truth.}
    \label{fig:huang_comparison}
\end{figure}

\smallskip
\noindent \textbf{{\cgf Qualitative} comparison to \textcite{huang2017shapesynth}.}
To complement our comparison to \cite{huang2017shapesynth} from \Cref{subsec:exp_reconstruction} we show a qualitative comparison of chairs from our dataset. \Cref{fig:huang_comparison} shows the reconstructions using both suggested networks as well as the resulting Chamfer Distances.

\smallskip
\noindent \textbf{Expressiveness comparison to other methods.}
For both StructureNet~\cite{mo2019structurenet} and ShapeAssembly~\cite{shapeAssembly} we use a point cloud encoder~\cite{qi2017pointnetplusplus} which was trained on PartNet~\cite{mo2019partnet} to map point clouds to the latent space. For ShapeAssembly we use the authors' encoder and for StructureNet we train the encoder ourselves. During inference, we use the decoders released by the authors to recover shapes from point clouds sampled from COSEG shapes~\cite{wang12aca}.
We use the COSEG~\cite{wang12aca} chair and vase shape sets of 400 chairs and 300 vases. Neither we nor the baseline methods have been trained on this dataset, so we find it a fair test for the expressiveness of each method. 
From each shape, we create point clouds with 1,500 points that are sampled using Farthest Point Sampling~\cite{eldar1997FPS} and an additional 800 randomly sampled points. \Cref{tab:baseline} shows the average Chamfer Distance for the chair and vase sets from COSEG for different reconstruction methods. We observe that our system produces more accurate reconstructions than the baselines.

\begin{figure}[t!]
    \centering
    \includegraphics[width=0.99\linewidth]{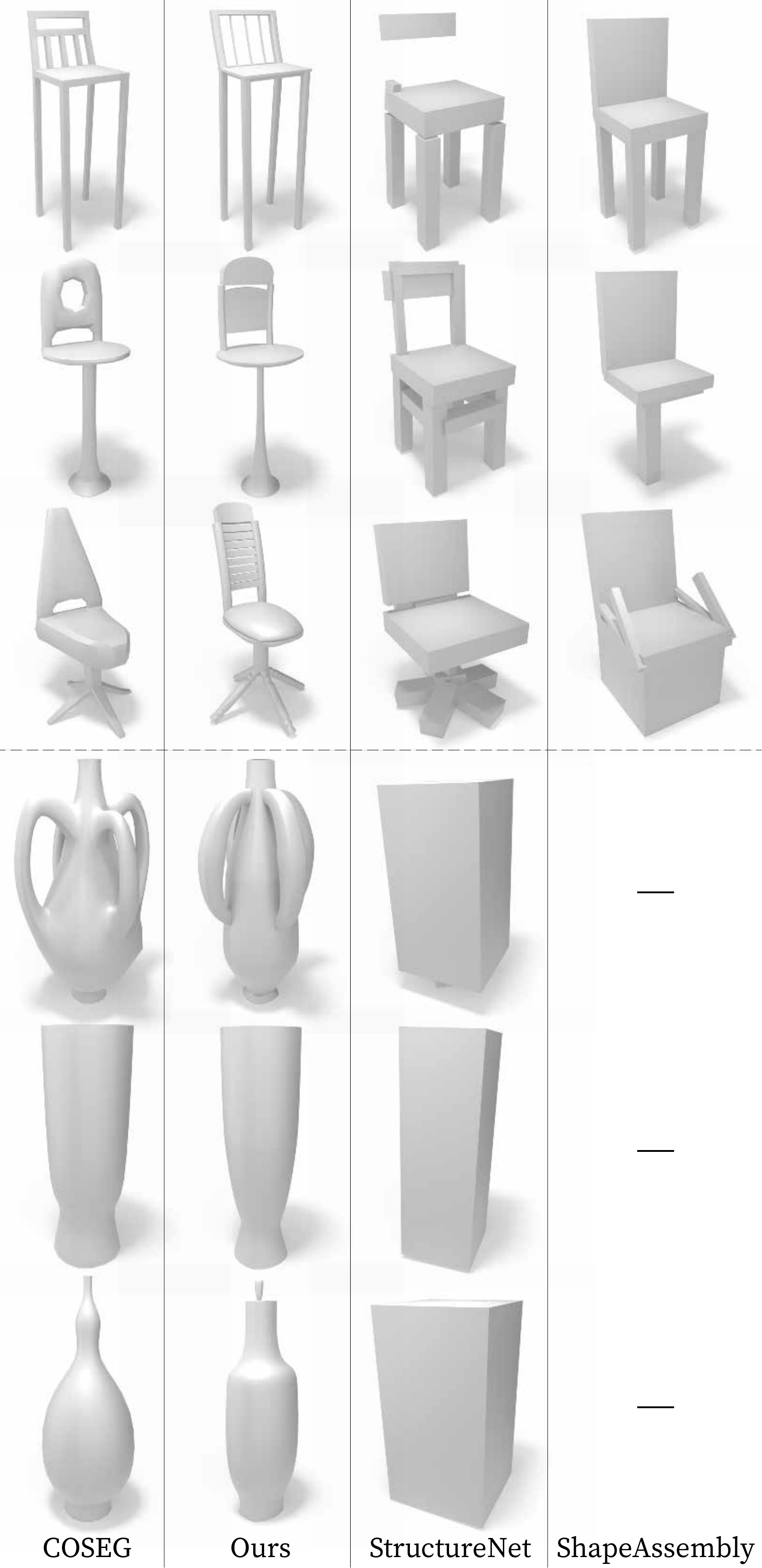}
    \caption{\textbf{Expressiveness comparison to other methods.} We show the reconstruction results of \ourmethod{}, StructureNet~\cite{mo2019structurenet}, and ShapeAssembly~\cite{shapeAssembly} on point cloud inputs generated from COSEG~\cite{wang12aca} samples. The compared methods yield crude approximations of the input geometry and may output a shape with detached parts. In contrast, \ourmethod{}'s predictions are much closer to the original shape while also being structurally sound.}
    \label{fig:other_methods}
\end{figure}

In \Cref{fig:other_methods} we present qualitative results that complement \Cref{tab:baseline}. The visual comparison shows the generalization power of \ourmethod{} while also explaining the drastic decrease in average Chamfer Distance compared to
both StructureNet~\cite{mo2019structurenet} and ShapeAssembly~\cite{shapeAssembly}.

\begin{table}[t]
\centering
\small
\begin{tabular}{ l c c } 
\toprule
Method / Category & Chairs $\downarrow$ & Vases $\downarrow$ \\
\midrule
StructureNet~\cite{mo2019structurenet} & 0.0203 & 0.0215 \\
ShapeAssembly~\cite{shapeAssembly} & 0.0134 & - \\
\ourmethod{} (ours) & \bf{0.0062} & \bf{0.0112}\\
\bottomrule
\end{tabular}
\caption{\textbf{Expressiveness comparison to other methods.} \ourmethod{} achieves significantly lower Chamfer Distance on both the evaluated domains of chairs and vases and demonstrates the importance of using expressive shape parts.
}
\label{tab:baseline}
\end{table}

\smallskip
\noindent \textbf{Reconstruction from point cloud \vs sketch.} We study the reconstruction performance of both point cloud and sketch inputs on our test sets for all three domains. Note that the calculation of the average Chamfer Distance for the sketch input type was done using only the camera angles that were seen during training.
\par
\begin{figure}[t!]
    \centering
    \includegraphics[width=0.95\linewidth]{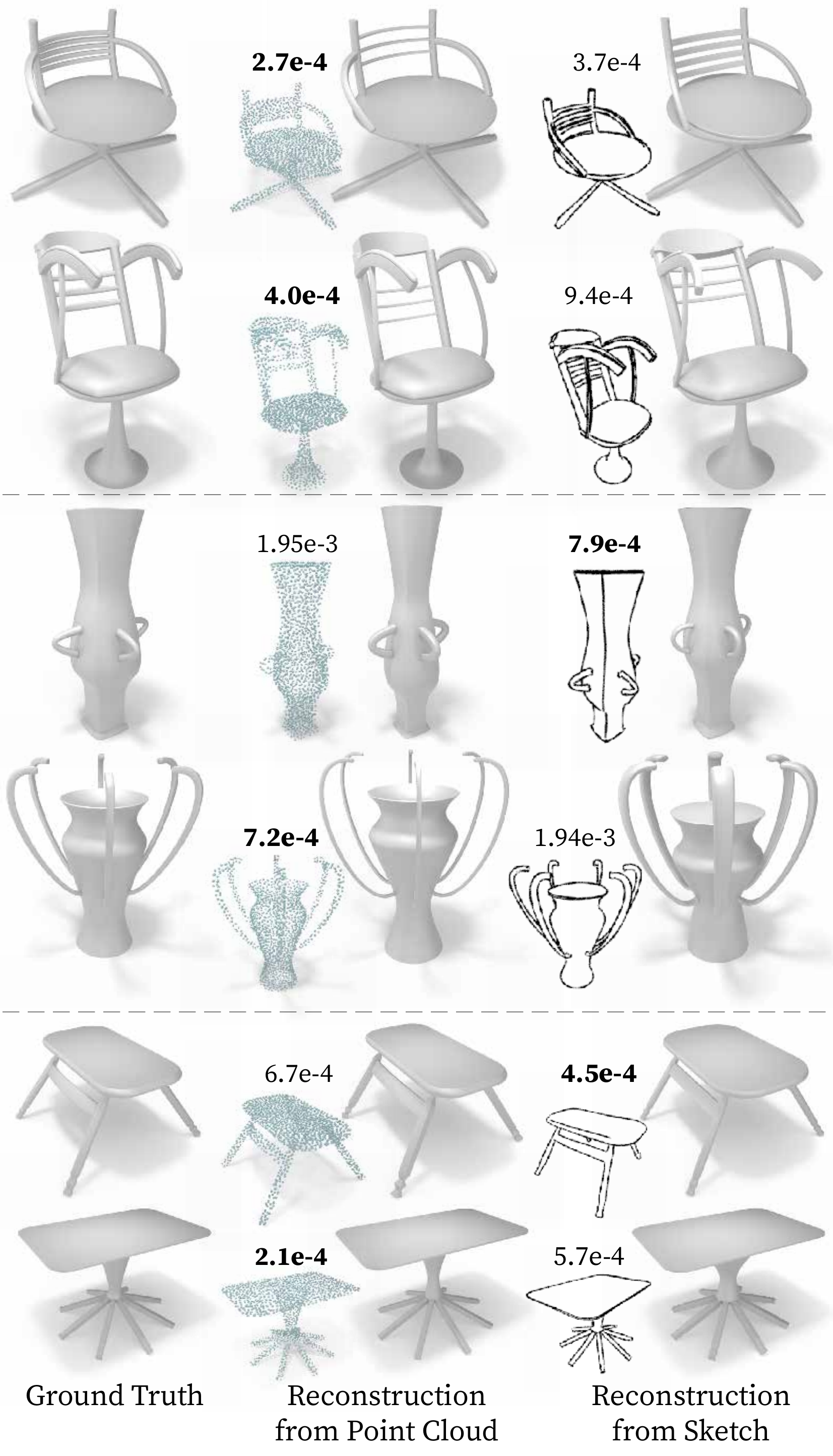}
    \caption{\textbf{Reconstruction from point cloud \vs sketch.} We compare the reconstruction results of \ourmethod{} from point clouds and sketches. The Chamfer Distance for each prediction is noted near the reconstruction result. In most cases, point cloud reconstruction yields a better result. However, sketch reconstruction may sometimes perform better (see the first vase and the first table).}
    \label{fig:pc_sketch}
\end{figure}

In \Cref{tab:pc_sketch} we see that, overall, reconstruction from point clouds generally performs better compared to reconstruction from sketch inputs. For example, \Cref{tab:pc_sketch} shows that for the chair domain, the average Chamfer Distance when reconstructing from point clouds is 2.5 times better compared to reconstruction from sketches. The vase and table domains show similar results. We attribute this, in part, to the fact that the scale of the shape cannot be effectively captured from a single image, while a point cloud inherently encodes the scale of the shape. The qualitative results for this experiment are shown in \Cref{fig:pc_sketch}.

\begin{table}[t!]
\small
\centering
\setlength{\tabcolsep}{3pt}
\begin{tabular}{l c c}
\toprule
Dataset & Point Cloud & Sketch \\
\midrule
Chair & 0.00040 & 0.00106 \\
Vase & 0.00264 & 0.00488 \\
Table & 0.00044 & 0.00250 \\
\reva{Ceiling Lamp} & \reva{0.00411} & \reva{0.01403} \\
\bottomrule
\end{tabular}
\caption{\textbf{Reconstruction from point cloud vs. sketch.} Showing the average Chamfer Distance for the reconstruction task on our test set. We compare the reconstruction results of the point cloud and sketch inputs. Note that when calculating the average Chamfer Distance for sketch input, we only consider sketches from camera angles that we trained on. Point cloud reconstruction performs better in all three domains.
}
\label{tab:pc_sketch}
\end{table}

\subsection{Shape Editing} \label{subsec:sup_shape_editing}

In this section, we complement the shape interpolation experiment \Cref{subsec:exp_edits} from the main paper and demonstrate interpolation over a selected set of parameters.

\begin{figure}[t!]
    \centering
    \includegraphics[width=0.99\linewidth]{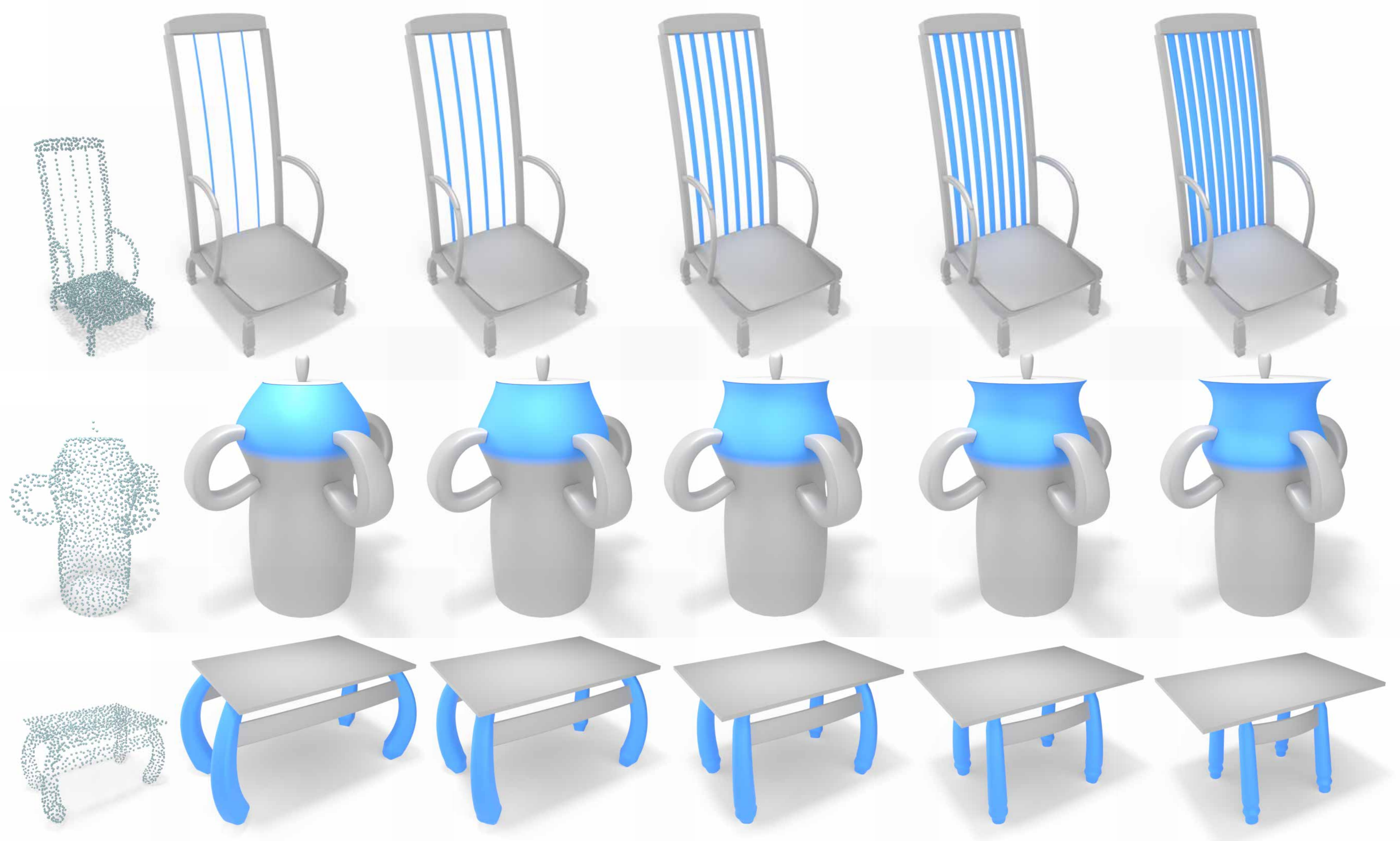}
    \caption{\textbf{Selective editing.} \ourmethod{} enables controlled editing over a particular attribute, while accurately adapting the remaining parts of the shape. Note the vase handles remain connected after each gradual edit.}
    \label{fig:selective_interpolate}
\end{figure}

\smallskip
\noindent \textbf{Selective Shape Interpolation.}
In contrast to interpolating the entire parameter set, ~\Cref{fig:selective_interpolate} shows interpolation over selected geometric features for $\alpha \in [0,1]$ in 0.25 increments.

\subsection{Robustness} \label{subsec:sup_exp_robustness}

In this section, we evaluate our method's robustness to slightly deformed inputs. Specifically, we both qualitatively and quantitatively evaluate the performance of \ourmethod{} on the following deformation: sketches of simplified shapes and point clouds sampled from simplified shapes, sketches rendered from novel angles, random point clouds with decreasing number of points, and point clouds with added Gaussian noise.

\smallskip
\noindent \textbf{Reconstruction from simplified shapes.}
Simplification reduces the number of polygons and, therefore, the level of detail in the shape. In this experiment, we show that our system is able to reconstruct shapes from both point cloud and sketch inputs that are produced from simplified meshes. We compare the non-simplified reconstruction where we have approximately 25K faces (this number varies depending on the visible parts in the shape) to simplified versions where the average number of faces is decreased by a factor of $10\times$ ($\sim$2K faces) and a factor of $100\times$ ($\sim$120 faces).

\begin{figure}[t!]
    \centering
    \includegraphics[width=0.99\linewidth]{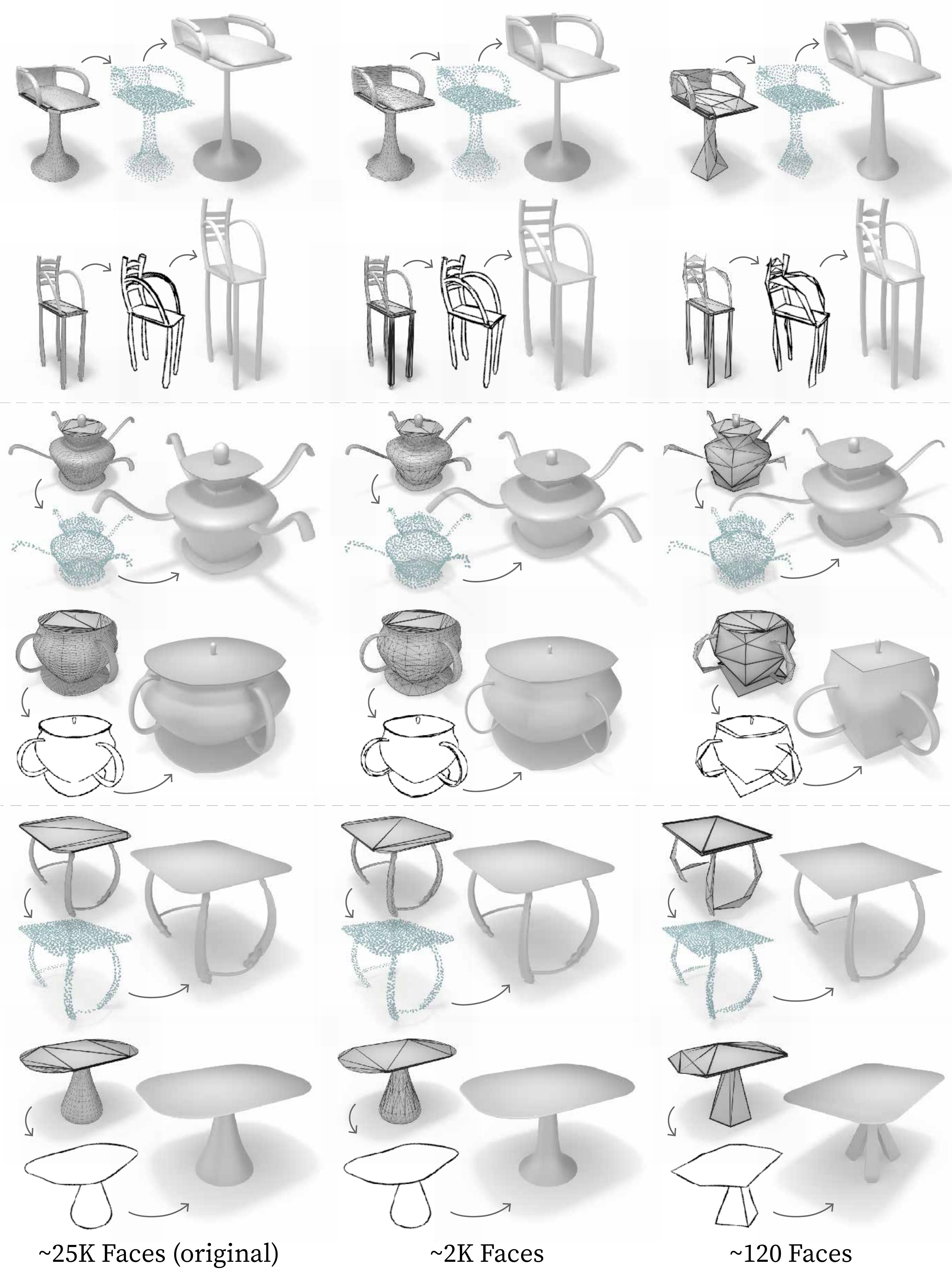}
    \caption{\textbf{Reconstruction from simplified shapes.} We perturb the test set via mesh simplification and show the reconstruction for point clouds and sketches. From left to right: the original shapes with no simplification ($\sim$25K faces), simplification factor of $\times 10$  ($\sim$2K faces), followed by a simplification factor of $\times 100$ ($\sim$120 faces). We observe that \ourmethod{} can still accurately recover the original shape, even when introducing modifications from simplified meshes. Noticeable artifacts appear for the chair reconstruction from the point cloud, particularly the leg of the chair lacks a solid base. Another example is the vase reconstruction from the sketch, where the handles are less accurate, and perhaps understandably, the vase assumes a square appearance. Also, notice that the leg of the second table is simplified in such a way that it is later confused with a swivel leg with three split legs that stem from the base of the tabletop.
    }
    \label{fig:simplified_meshes}
\end{figure}

\Cref{tab:simplified_meshes} shows the average Chamfer Distance for each domain. For all the domains, both the point cloud and the sketch reconstruction experience only a slight increase in the average Chamfer Distance for the lower simplification factor of $\times 10$ ($\sim$2K faces). However, a drastic increase in the average Chamfer Distance is registered for both input types at the higher simplification factor $\times 100$ ($\sim$120 faces). We show the qualitative results for this experiment in \Cref{fig:simplified_meshes}.

\begin{table}[b]
\centering
\small
\setlength{\tabcolsep}{2pt}
\begin{tabular}{@{ } l l c c c @{ }}
\toprule
Dataset & Input & $\sim$25K Faces $\downarrow$ & $\sim$2K Faces $\downarrow$ & $\sim$120 Faces $\downarrow$ \\
\midrule
\multirow{2}{2em}{Chair} & PC & 0.00040 & 0.00041 & 0.00133 \\
& Sketch & 0.00106 & 0.00117 & 0.00227 \\
\midrule
\multirow{2}{2em}{Vase} & PC & 0.00264 & 0.00280 & 0.00613 \\
& Sketch & 0.00488 & 0.00517 & 0.01160 \\
\midrule
\multirow{2}{2em}{Table} & PC & 0.00043 & 0.00044 & 0.00149 \\
& Sketch & 0.00222 & 0.00308 & 0.00520 \\
\bottomrule
\end{tabular}
\caption{\textbf{Reconstruction from simplified shapes.} Comparing the average Chamfer Distance on our test set at different mesh simplification factors. The averages only slightly increase for a simplification factor of $10\times$ ($\sim$2K faces) compared to the original mesh ($\sim$25K faces) and increase more drastically for a high simplification factor of $100\times$ ($\sim$120 faces).
}
\label{tab:simplified_meshes}
\end{table}

\smallskip
\noindent \textbf{Reconstruction from randomly sampled point clouds.}
Our point clouds during training and testing contain 1,500 points sampled using Farthest Point Sampling~\cite{eldar1997FPS} and an additional 800 randomly sampled points. In this experiment, we test the robustness of reconstruction performance when the system is given a decreasing number of exclusively randomly sampled points. For each shape in our test set, we randomly sample point clouds with 2,048, 1,024, and 512 points. We then perform reconstructions from these point clouds and compute the average Chamfer Distance between the reconstructed and ground-truth shapes.

\begin{figure}[t]
    \centering
    \includegraphics[width=0.99\linewidth]{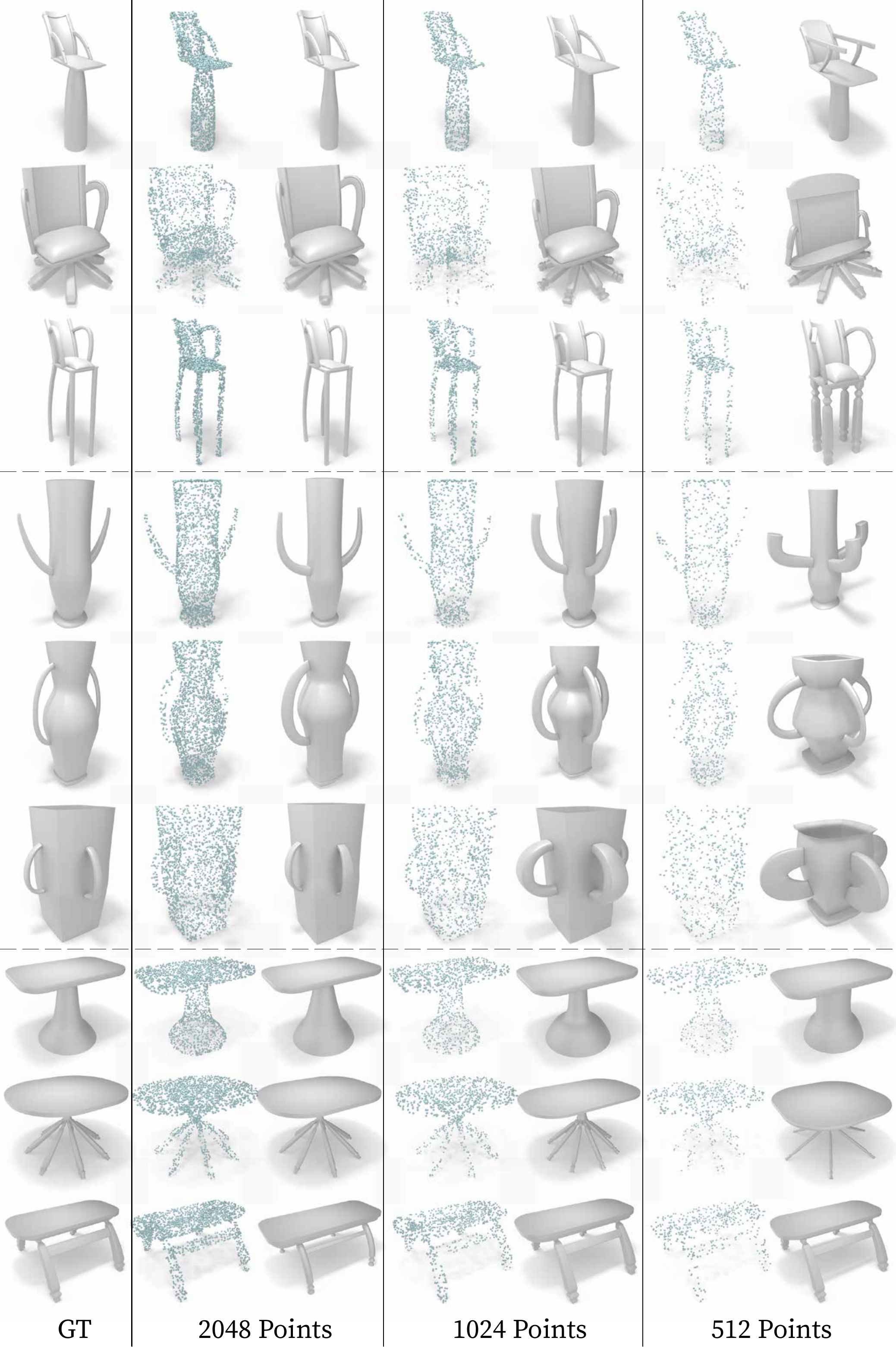}
    \caption{\textbf{Reconstruction from randomly sampled point clouds.} We use random sampling with a progressively decreasing number of points. At 2048 points, \ourmethod{} produces accurate reconstructions. A minor degradation in the level of detail appears for 1024 points, \eg, the second vase's body is slightly different compared to the ground truth. Lastly, at only 512 points, more considerable reconstruction inaccuracies start to appear. An example of this is seen in the first and last vases, where the body and the handles are less accurate. Another example is the second chair, where the seat height, top rail, and armrests are inaccurate. 
    }
    \label{fig:random_pc}
\end{figure}

\Cref{tab:random_pc} shows {\new the average Chamfer Distance when reconstructing from point clouds with a decreasing number of randomly sampled points. We note a steady increase in} the average Chamfer Distance between the reconstructed shapes and the ground-truth shapes. However, the {\new reconstructed shapes are still visually similar to the ground-truth shapes, as evident from}~\Cref{fig:random_pc}. In the last decrement, with only 512 points, our system produces shapes that resemble the overall structure of the ground-truth shape, however, some details are captured with degraded accuracy.

\begin{table}[t]
\centering
\begin{tabular}{l c c c}
\toprule
Dataset & 2048 & 1024 & 512 \\
\midrule
Chair & 0.00062 & 0.00311 & 0.01972 \\
Vase & 0.00361 & 0.00770 & 0.01914 \\
Table & 0.00053 & 0.00135 & 0.00536 \\
\bottomrule
\end{tabular}
\caption{\textbf{Reconstruction from randomly sampled point clouds.} We report the average Chamfer Distance between reconstructed and ground-truth shapes on our test set when reconstructing from randomly sampled point clouds with a varying number of points. Using 2048 points only slightly increases the average Chamfer Distances for all domains when compared to the 1500 Farthest Point Sampling in addition to 800 randomly sampled points that were used in \Cref{tab:pc_sketch}. Decreasing the number of points further to 1024, and then to 512, has a significant effect on the results. However, as shown in \Cref{fig:random_pc}, the reconstructions are performing well at 1024 points and suffer inaccuracies mostly at 512 points.
}
\label{tab:random_pc}
\end{table}

\smallskip
\noindent \textbf{Reconstruction from point clouds with Gaussian noise.} To further examine the robustness of our method we add varying levels of Gaussian noise to the point clouds in our test set and evaluate the reconstruction both qualitatively and quantitatively. We consider Gaussian noise with 0 mean and a progressively increasing standard deviation of 0.5\%, 1.0\%, and 2.5\% of the overall size of the {\new normalized} shape, corresponding to $\sigma = 0.01, 0.02$, and $0.05$.
\par
\begin{figure*}[t!]
    \centering
    \includegraphics[width=0.80\textwidth]{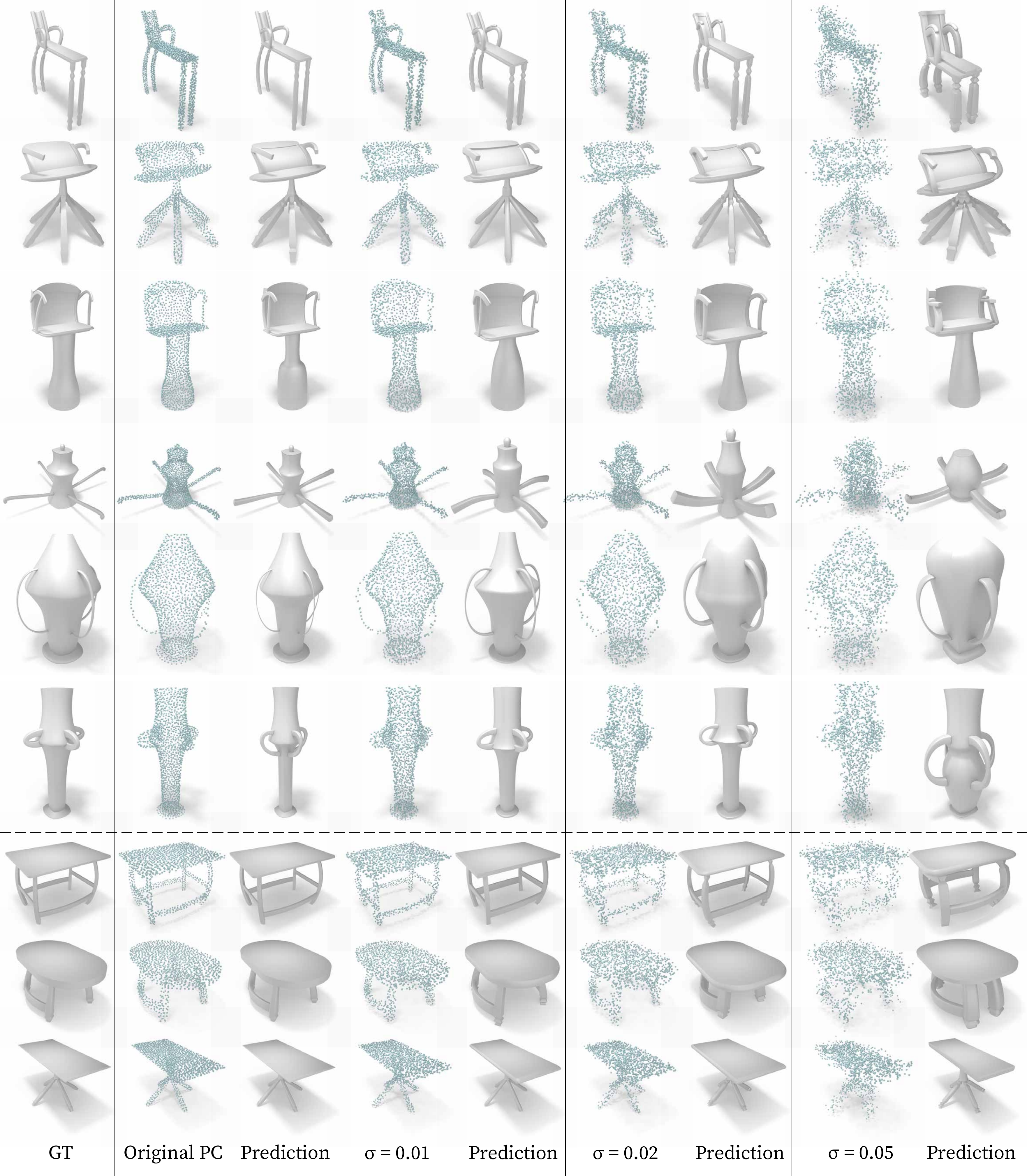}
    \caption{\textbf{Reconstruction from point clouds with Gaussian noise.} Qualitative comparison of the reconstruction from point clouds with varying levels of Gaussian noise with 0 mean and standard deviations of $\sigma = 0.01, 0.02$, and $0.05$.
    {\new In most cases}, we can see that our method produces a rather accurate prediction {\new up to} $\sigma = 0.02$, but the reconstructed shapes for $\sigma = 0.05$ have less resemblance to the {\new ground-truth shapes}. The first chair and the first table are examples that show an expected property where the shapes reconstructed from the noisy point cloud tend to appear thicker overall.}
    \label{fig:gaussian}
\end{figure*}

~\Cref{tab:gaussian} shows that our method remains somewhat resilient up to $\sigma = 0.02$, and begins to struggle when the noise level increases to $\sigma = 0.05$. The qualitative analysis in \Cref{fig:gaussian} is consistent with this result.

\begin{table}[t]
\centering
\begin{tabular}{@{ } l c c c c @{ }}
\toprule
Dataset & No Noise & $\sigma = 0.01$ & $\sigma = 0.02$ & $\sigma = 0.05$ \\
\midrule
Chair & 0.00040 & 0.0007 & 0.00125 & 0.00534 \\
Vase & 0.00264 & 0.00346 & 0.00492 & 0.01116 \\
Table & 0.00044 & 0.00076 & 0.00155 & 0.00577 \\
\bottomrule
\end{tabular}
\vspace{0.1cm}
\caption{\textbf{Reconstruction from point clouds with Gaussian noise.} We detail the average Chamfer Distance on our test set when reconstructing from point clouds with various noise levels. %
Our method is resilient to Gaussian noise when the standard deviation is 0.01, and experiences inaccuracies when more noise is added.
}
\label{tab:gaussian}
\end{table}

\begin{figure}[t!]
    \centering
    \includegraphics[width=0.99\linewidth]{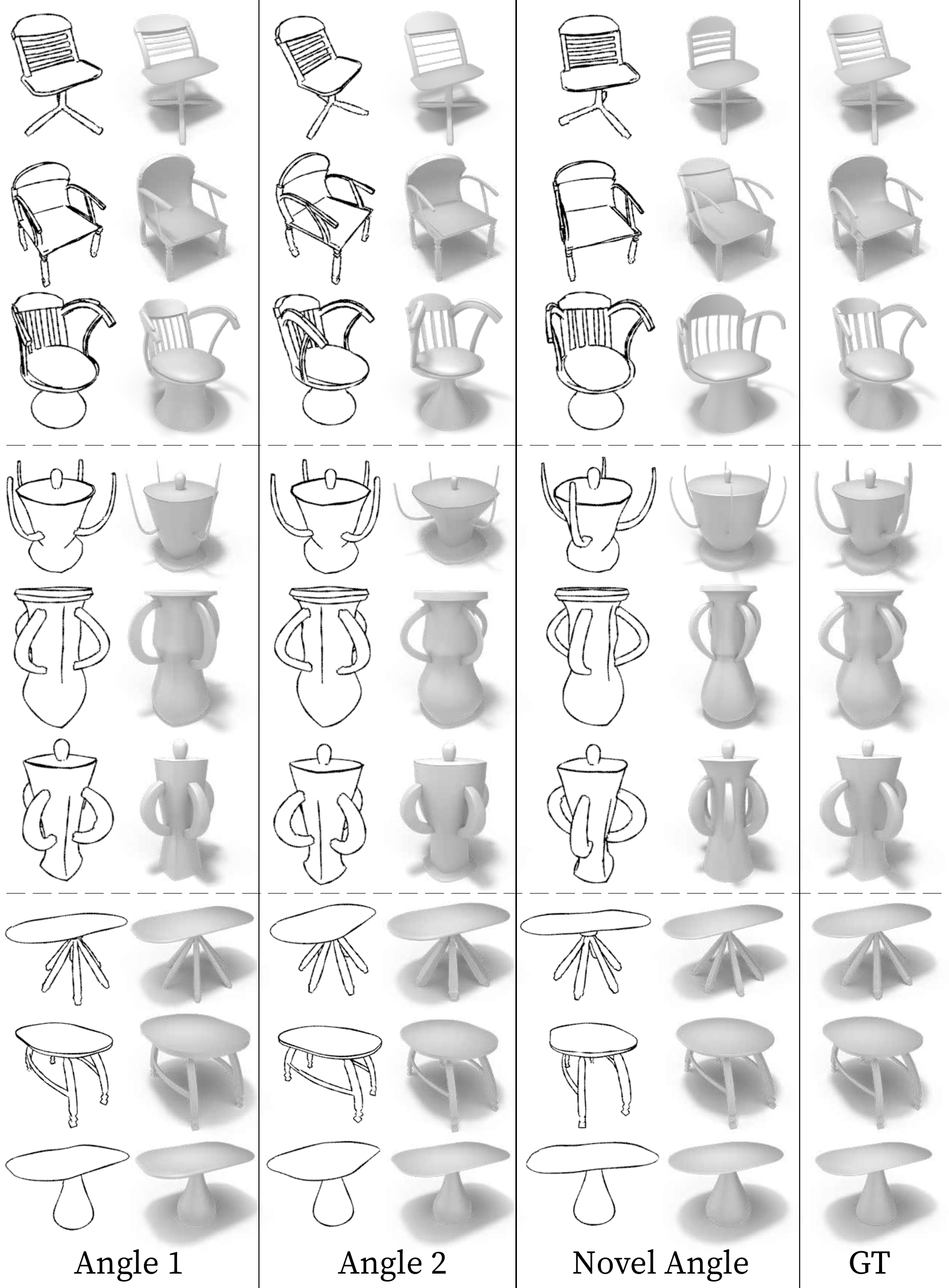}
    \caption{\textbf{Reconstruction from sketches of various camera angles.} Comparison between reconstruction from sketches rendered at different camera angles. The novel angle is an angle that was never seen during training.  The two first angles perform similarly with accurate results, while the novel angle produces less accurate results. For example, in the second chair, the backrest is slanted backward for the reconstruction from the novel angle. However, the ground-truth shape and the predictions from the first two angles show a forward-slanted backrest. Another example is the second vase, where the body is noticeably thinner.}
    \label{fig:sketch_angles}
\end{figure}

\smallskip
\noindent \textbf{Reconstruction from sketches of various camera angles.} 
We compare the reconstruction from sketches rendered from three different angles. During training, our system sees two of the angles while we leave the third \textit{novel angle} unseen. {\new The camera elevation is identical for all angles at 55\textdegree, and the azimuth is 35\textdegree, 55\textdegree, and 15\textdegree, for angle 1, angle 2, and the novel angle, respectively.} For each shape in our test set, we render sketches from all three angles and perform reconstructions from these sketch images. ~\Cref{tab:sketch_angles} shows that the average Chamfer Distance between {\new the} reconstructed and ground-truth shapes is similar when we reconstruct from the two sketch angles we trained on. However, reconstructing from the novel angle results in a larger Chamfer Distance. ~\Cref{fig:sketch_angles} shows that the shapes reconstructed from the novel angle are still visually close to the ground truth.
{\new We refer the user to \Cref{subsec:sup_failure_cases} and specifically \Cref{fig:failure_case_sketch_angles} where we discuss a common failure case when reconstructing shapes from the novel sketch angle.}

\begin{table}[t!]
\centering
\begin{tabular}{l c c c}
\toprule
Dataset & Angle 1 & Angle 2 & Novel angle \\
\midrule
Chair & 0.00108 & 0.00103 & 0.00807 \\
Vase & 0.00484 & 0.00491 & 0.00979 \\
Table & 0.00246 & 0.00254 & 0.00853 \\
\bottomrule
\end{tabular}
\caption{\textbf{Reconstruction from sketches of various camera angles.} We provide the average Chamfer Distance between reconstructed and ground-truth shapes in our test set for sketch inputs, comparing the result on sketches generated from the two camera angles that were seen during training with a novel angle unseen during training. The first two angles perform similarly. However, the average Chamfer Distance for the novel angle is increased significantly. Another observation is that the increase in the novel angle is not as pronounced for the vase domain since {\new the camera elevation is identical in all tested angles and} the vases tend to {\new inhabit} rotational symmetry.
}
\label{tab:sketch_angles}
\end{table}

\smallskip
\noindent \textbf{Structural stability.}
{\new During the dataset generation we take measures to generate structurally stable samples. For instance, a vase could have a base that is too narrow and will make it tip over under the weight of its handle. However, during inference time, we do not disqualify or enforce any stability-inducing modification on our predictions.

\begin{table}[t]
\centering
\small
\begin{tabular}{c c c c}
\toprule
Dataset & \thead{Stable GT \\ Samples} & \thead{Stable PC \\ Predictions} & \thead{Stable Sketch \\ Predictions} \\
\midrule
Chair & 91.01\% & 90.28\% & 91.54\% \\
Vase & 100.00\% & 99.79\% & 99.57\% \\
Table & 99.84\% & 96.33\% & 97.24\% \\
\bottomrule
\end{tabular}
\caption{
{\textbf {Structural stability.}} We show the percentage of ground truth and predicted samples from our test set that are structurally stable according to our \textit{stability} metric. We show a high percentage of ground truth sample stability and matching high percentages for our predicted shapes.
}
\label{tab:stability}
\end{table}

This means that our network may produce unstable predictions. To ensure that the network successfully learns the distribution of stable shapes, we define a binary \textit{stability} metric. We consider a shape to be \textit{stable} if all its parts are attached, and it remains standing in a physics simulation after a drop onto a flat plane from a height of 5\% of the shape's height. In~\Cref{tab:stability}, we can see that \ourmethod{}'s ground-truth shapes are mostly stable, this is unsurprising, however, the predictions that our system generates are on par with the ground-truth data in terms of their stability which is another indication that our training process was successful.}

\subsection{Ablation Study} \label{subsec:sup_exp_ablation}

In this section, we discuss another ablation study related to our shape recovery network, you can find the other ablation study in section \Cref{sec:exp} of the main paper.

\smallskip
\noindent \textbf{Joint training \vs separate training.} In this experiment we wish to find out whether training separately improves the average Chamfer Distance for recovered shapes. We train the network as two separate networks, one for point cloud inputs and another for sketch inputs. To begin with, the encoders (point cloud encoder and sketch encoder) are already separated, but we have to duplicate the decoders in such a way that both networks now have their own decoding network. We note that this effectively doubles the number of weights in the decoding part of each separate network.

\begin{table}[t]
\centering
\begin{tabular}{l c c}
\toprule
Method & Point Cloud $\downarrow$ & Sketch $\downarrow$ \\
\midrule
Jointly & 0.00040 & 0.00106 \\
Separately & \textbf{0.00037} & \textbf{0.00098} \\
\bottomrule
\end{tabular}
\caption{\textbf{Joint training \vs separate training.} We report the average Chamfer Distance for two training methods. Jointly: we train the network as described in \Cref{fig:overview} and in \Cref{sec:method}. Separately: we train on point cloud inputs and sketch inputs separately while keeping the rest of the network without a change. We note that training separately allows the decoders to allocate all their available weights to a single input type. However, separate training requires two sets of decoders, in contrast to one set used during the joint training.
}
\label{tab:jointly}
\end{table}

In \Cref{tab:jointly}, we show the average Chamfer Distances for point cloud and sketch reconstruction for both of these methods. Training separately yielded a minor improvement in both point cloud and sketch reconstructions but at the cost of additional complexity.

\subsection{Failure Cases} \label{subsec:sup_failure_cases}

\begin{figure}[t!]
    \centering
    \includegraphics[width=0.99\linewidth]{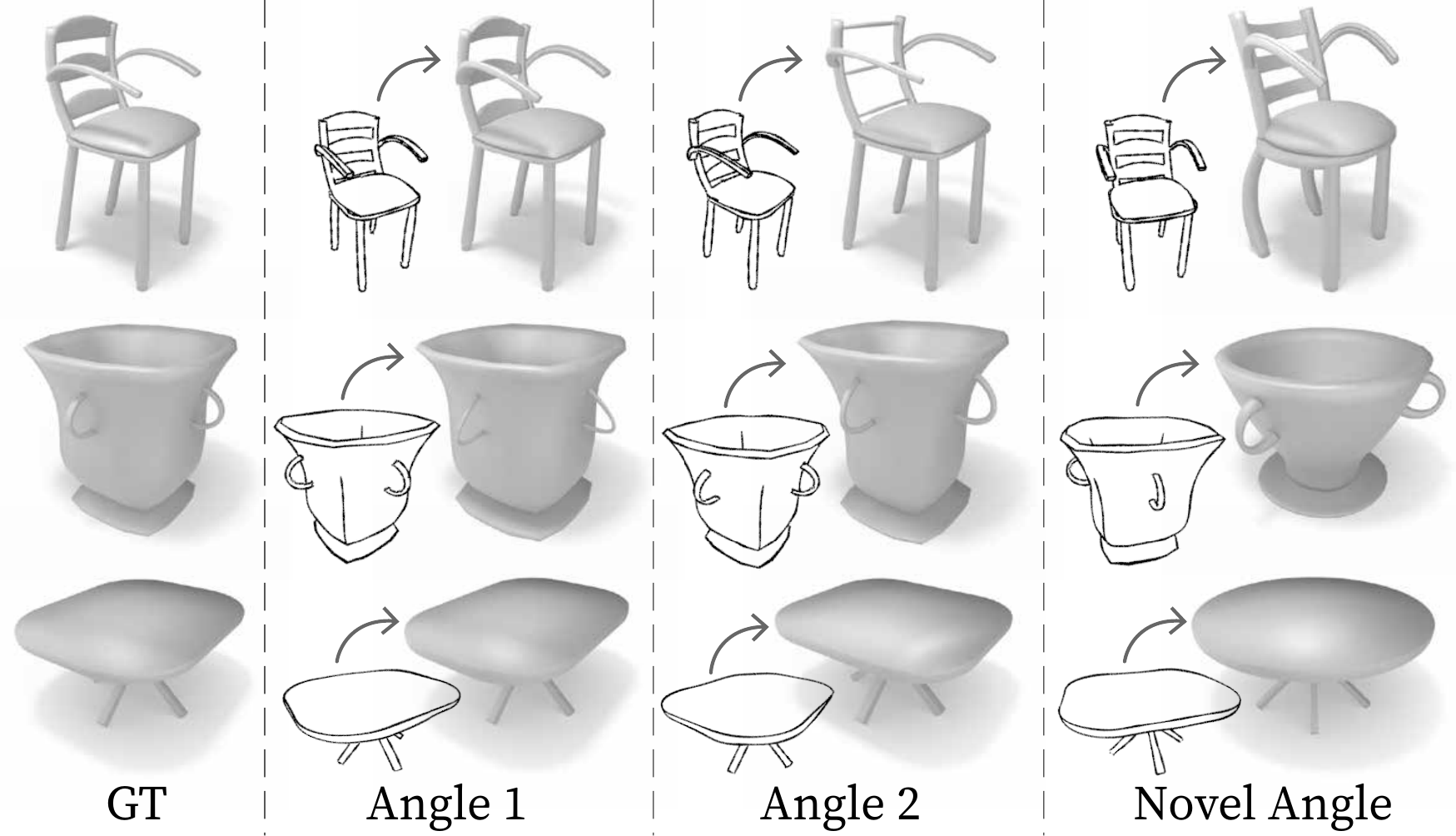}
    \caption{\new{\textbf{Rounded reconstructions for the novel sketch angle.} We show a limitation of \ourmethod{} where the reconstructions from sketches that were drawn from the novel camera angle often exhibit a tendency for rotational symmetry that causes the increase in Chamfer Distance \reva{as shown} in \Cref{tab:sketch_angles}.}}
    \label{fig:failure_case_sketch_angles}
\end{figure}

\smallskip
\noindent \textbf{Rounded reconstructions for the novel sketch angle.}
In \Cref{tab:sketch_angles} we discuss a significant increase in Chamfer Distance when reconstructing shapes from the novel sketch view compared to sketch view angles on which we trained. While analyzing this increase we recognized a pattern in the shapes that were reconstructed from the novel sketch view where in many cases the shapes took a more rounded shape. We exemplify this tendency for rotational symmetry in \Cref{fig:failure_case_sketch_angles}.

\begin{figure}[t!]
    \centering
    \includegraphics[width=0.99\linewidth]{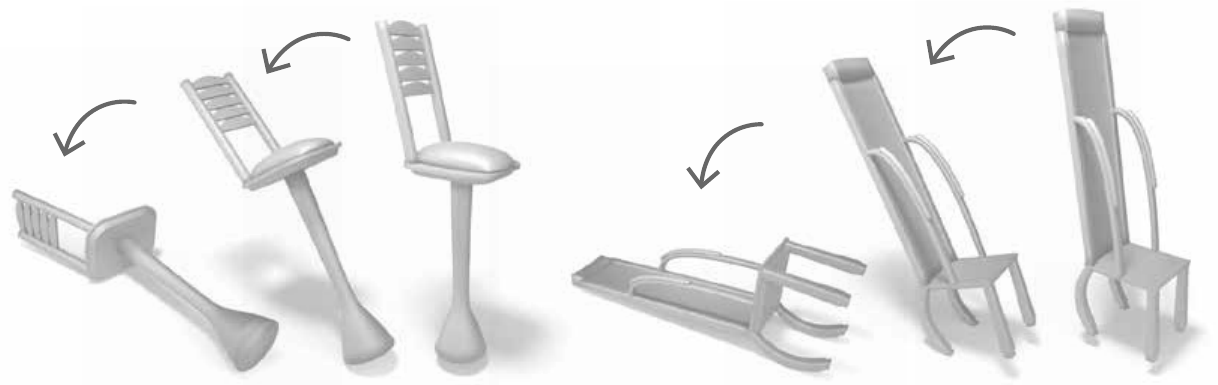}
    \caption{\new{\textbf{Failed shapes from the physical simulation.} Examples of chairs from our test set that failed in our physics simulation when dropped from a height of 5\% of the shape's height. The chairs tend to be tall and have a narrow base.}}
    \label{fig:failure_case_simulation}
\end{figure}

\smallskip
\noindent \textbf{Failed shapes from the physical simulation.}
{\new In \Cref{sec:sup_experiments} when we discuss the structural stability metric, we explain that we take heuristic measures to ensure some stability on the samples in our dataset. In \Cref{fig:failure_case_simulation} we show examples from our test set that fell in our physics simulation. This means that our datasets are not perfect in that regard, however, as \Cref{tab:stability} suggests, the predictions, where we do not employ any stability-inducing heuristics, do not incur a higher percentage of unstable samples.}

\section{Additional Implementation Details} \label{sec:sup_impl}

\subsection{Dataset} \label{subsec:sup_dataset}

\smallskip
\noindent \textbf{{\cgf Part existence label}.}
In our program, we include support for binary properties that control the visibility of certain parts of the shape, \eg, a chair may or may not include armrests. To induce a probability distribution function on all the parameters in our human-interpretable parameter space, we add a \textit{part existence label} to parameters that belong to a switchable part of the shape. This allows the use of a subset of the parameters when predicting a shape, allowing the network to predict that a certain parameter has no effect on the final shape.

Continuing the armrests example, all the parameters that control the shape of the armrests may have no effect on the final chair shape if the binary parameter controlling the visibility of the armrests is set to false, and hence, the final chair shape will not include armrests. Consequently, these parameters will be given the option to be set to our part existence label. A pre-processing stage on our datasets assigns any such parameter to the part existence label if the part it controls is not visible in the shape. This step is done based on a set of \textit{visibility rules} that are \textit{inferred automatically} {\cgf by recursively analyzing} the shape program's node graph.

When we compare classification and regression in our ablation study in \Cref{subsec:sup_exp_ablation}, we incorporate the part existence label into the regression to make the comparison fair. To achieve that, each decoder of a continuous parameter that controls a part of the shape that can be invisible will output two elements instead of one. The first element will be the predicted value of the parameter, while the second element predicts the existence label. Whenever the prediction of the second element exceeds 0.5 we say that the part is not visible in the final shape, and the actual predicted value will be the existence label. During loss calculation, we accumulate the Mean Square Error of both elements. If the existence label was predicted (the second element is greater than 0.5) then we avoid collecting the regression loss of the first element (which predicts the value of the parameter).

\smallskip
\noindent \textbf{Dataset preprocessing.} In the data generation step, we create for each sample an OBJ file containing the shape, the sketches rendered from three camera views, and a YAML file containing the labels of each parameter. Prior to training, we execute a preprocessing step where we prepare for each generated sample, two point clouds, and one YAML file which holds the normalized values of the parameters.

We begin by sampling 1500 points of the generated object mesh using Farthest Point Sampling~\cite{eldar1997FPS}, then another point cloud using random sampling which will be used for augmentation. When we retrieve a sample we take all the Farthest Point Sampled points and randomly pick 800 points from the 1,500 randomly sampled points. 

Finally, we convert the shape's YAML file to its normalized form, according to the following conditions: 1) integer parameters will start from 0,  2) float (and vector) parameters will be in the range of [0.0, 1.0] mapping the original values to that range is done linearly, 3) any parameter which is not visible in the final shape will be set to the \textit{existence label} which is implemented with the value -1.0. The idea here is to avoid forcing the network to decide on a label for parameters that are not visible.

\smallskip
\noindent \textbf{Recipe file.}
To allow for an easy dataset generation according to the needs of the user, we created the \textit{recipe} configuration file which contains the instructions to create the dataset.
\par
The recipe file describes the minimum and maximum allowed values for each parameter. Continuous parameters also include the sampling granularity which is used during the data generation step to split the allowed range uniformly with a number of samples matching the user-specified sampling value.

\subsection{Network Architecture} \label{subsec:sup_network}

\smallskip
\noindent \textbf{Point cloud encoder.} We base our point cloud encoder on the four-layer classification architecture of DGCNN~\cite{wang2019dynamic}. Each layer is comprised of two parts, k-NN and then \textit{EdgeConv}. k-NN finds the closest k points of each point. In the beginning, the closeness is determined in terms of physical distances between the points, but in the next layers, the closeness is determined in the graph-feature space. In our testing, we use {\new the default} $k=20$.
\par
For each point $x_{i}$, we take all k closest points $x_{i,j}$ to it, and EdgeConv will have $k$ inputs, where each one is the concatenation of the features of $x_{i}$ and the features of $x_{i}-x_{i,j}$. The concatenated features are fed to a multi-layer perceptron with a matching input size, a chosen channel size, and no hidden layers. The output is then followed by a batch normalization and then a Leaky ReLU with a negative slope of 0.2. For EdgeConv will take the maximum between all the outputs. After EdgeConv is completed for all the points we continue to the next k-NN and EdgeConv pair.
\par
We choose the channel size in each of the four EdgeConv layers to be 16, 16, 32, and 64. The outputs from the four layers are aggregated together and enter another multi-layer perception, with no hidden layer and an output size of 64. Finally, max and average pooling are applied to that result and aggregated to form our embedding vector of size 128.

\smallskip
\noindent \textbf{Sketch encoder.} For the sketch encoder we base our architecture on VGG11~\cite{simonyan2015a_vgg} encoder. Our encoder assumes the following numbers of channels 
$$[8, M, 16, M, 32, 32, M, 64, 64, M, 64, 64, M]$$ where $M$ is a 2D max-pooling layer with kernel size 2 and stride 2. The numbers specify the number of channels in each 2D convolution layer, all have a kernel size of 3 and {\new a} padding {\new value} of 1, and each one is followed by a batch normalization and then ReLU. The result goes through {\new average} pooling and an additional linear layer which outputs an embedding vector of size 128.
\par
The input size to the encoder remains unchanged at $224 \times 224$ square images. In our tests we only use grayscale images of sketches, so we further optimize to only have a single channel for the input images.

\smallskip
\noindent \textbf{Decoding network.} As already stated, each decoder in the decoding phase is a multi-layer perception made out of three layers. The first and second layers are each followed by batch normalization, Leaky ReLU with a negative slope of 0.2 then dropout with a probability of 0.5. As shown in the System Overview~\Cref{fig:overview}, the input to the multi-layer perceptron is the embedding vector which, as stated, has a size of 128, the next two hidden layer{\new s'} sizes are 128 and 64. Finally, the output size is dictated by the number of labels of that parameter, in addition to the \textit{part existence label} when required as discussed earlier in \Cref{subsec:sup_dataset}.

\subsection{Training Scheme} \label{subsec:sup_training}
We train our network for 600 epochs, with an ADAM optimizer, an initial learning rate of 1e-2, and a scheduler with a step size of 20 and a gamma value of 0.9. Our batch size was 33. {\new Training on an Nvidia RTX 2080Ti card takes 55 hours and inference takes 20 milliseconds for a batch with a single sample.}

\end{document}